\documentclass[superscriptaddress,nofootinbib,preprintnumbers,twocolumn,10pt,aps,prd]{revtex4-2}
\usepackage{amsmath}
\usepackage{amssymb}
\usepackage{datetime}
\usepackage{graphicx}
\usepackage{hyperref}
\usepackage{cprotect}

\usepackage{blindtext}
\usepackage{hyperref}
\usepackage{amsmath,amssymb}
\usepackage{float}
\usepackage{microtype}
\usepackage{graphicx}
\usepackage{bm}
\usepackage{latexsym}
\usepackage{epsfig}
\usepackage{psfrag}
\usepackage{color}
\usepackage[dvipsnames]{xcolor}

\begin{document}


\preprint{KCL-2024-34}
\title{Refining Galactic primordial black hole evaporation constraints}

\author{Pedro De la Torre Luque}
\email{pedro.delatorre@uam.es}
\affiliation{
    Instituto de F\'isica Te\'orica, IFT UAM-CSIC,
    Departamento de F\'isica Te\'orica,
    Universidad Aut\'onoma de Madrid,
    ES-28049 Madrid,
    Spain
}

\author{Jordan Koechler}
\email{jordan.koechler@gmail.com}
\affiliation{
    Laboratoire de Physique Th\'eorique et Hautes Energies (LPTHE),
    UMR 7589 CNRS \& Sorbonne Universit\'e,
    4 Place Jussieu, F-75252,
    Paris, France
}

\author{Shyam Balaji}
\email{shyam.balaji@kcl.ac.uk}
\affiliation{
    Physics Department,
    King's College London,
    Strand, London, WC2R 2LS,
    United Kingdom
}

\date{\formatdate{\day}{\month}{\year}, \currenttime}

\begin{abstract}
    We revisit the role of primordial black holes (PBHs) as potential dark matter (DM) candidates, particularly focusing on light asteroid-mass PBHs. These PBHs are expected to emit particles through Hawking evaporation that can generate cosmic rays (CRs), eventually producing other secondary radiations through their propagation in the Milky Way, in addition to prompt emissions.
    Here, we perform a comprehensive analysis of CR signals resulting from PBH evaporation, incorporating the full CR transport to account for reacceleration and diffusion effects within the Milky Way. In particular, we revisit the $e^\pm$ flux produced by PBHs, using {\sc Voyager 1}, and study for the first time the diffuse X-ray emission from the up-scattering of Galactic ambient photons due to PBH-produced $e^\pm$ via the inverse Compton effect using {\sc Xmm-Newton} data, as well as the morphological information of the diffuse $511$~keV line measured by {\sc Integral/Spi}.
    In doing so, we provide leading constraints on the fraction of DM that can be in form of PBHs in a conservative way, whilst also testing how different assumptions on spin and mass distributions affect our conclusions.
    
\end{abstract}

\maketitle

\section{Introduction}
The continuous non-detection of definitive, non-gravitational signals from dark matter (DM) candidates, especially those within the weakly interacting massive particles category \cite{ParticleDataGroup:2022pth}, has sparked renewed interest in primordial black holes (PBHs) \cite{Carr:2020gox, Auffinger:2022khh,Escriva:2022duf}. While gravitational lensing significantly constrains the DM fraction that PBHs can account for at higher masses \cite{Carr:2020gox}, it is nearly impossible to detect PBHs with masses much below $10^{20}$ g due to finite-size source effects \cite{Smyth:2019whb,Niikura:2017zjd}. Nevertheless, black holes around this mass range are expected to emit intense non-thermal radiation through Hawking evaporation, providing an alternative avenue to constrain their contribution to DM \cite{Auffinger:2022khh}.

The temperature of black holes is inversely proportional to their mass, with $T \sim 0.1$ keV $(10^{20} \text{ g}/M_{\rm BH})$, meaning the radiation emitted for masses below $10^{20}$ g is anticipated in the X-ray to $\gamma$-ray spectrum. Furthermore, if $M_{\rm BH} \lesssim 10^{16}$ g, the evaporation process also produces positrons in abundance. These positrons, upon annihilation with ambient electrons, generate detectable $\gamma$-rays, particularly at an energy of $E_\gamma = m_e = 511$ keV \cite{MacGibbon:1991vc}. This finding has prompted efforts to determine the mass fraction of black holes in the vicinity of $M_{\rm BH} \sim 10^{16}$ g or potentially greater, by utilizing 511 keV observations \cite{MacGibbon:1991vc, Bambi:2008kx, Carr:2009jm,Keith:2021guq}. Notably, the recent study Ref.~\cite{DeRocco:2019fjq} established rather cautious constraints based on the 511 keV emissions detected from the center of our galaxy.

The 511 keV line is not the only signal though, with indirect searches for charged particle injection (that does not form positronium) from PBH evaporation like antiprotons, electrons, and positrons having been examined as well \cite{Barrau:1999sk,Carr:2016hva}. The major challenge with using these particles to constrain PBHs is their susceptibility to the Sun's influence at sub-GeV energies, which significantly supresses their flux when entering the heliosphere. On the other hand, low-energy physics is crucial because, the greater the mass of the PBH, the lower the energy of the emitted particles. Since the {\sc Voyager 1} probe has passed beyond the heliopause, it has detected low-energy electrons and positrons \cite{stone2013voyager,2016ApJ...831...18C} that may originate from PBH evaporation \cite{Boudaud:2018hqb} and is not affected by solar screening. 

Also, non-observation of Hawking radiation signatures from PBHs evaporation in the keV-MeV energy band probing the inner regions of the Milky Way \cite{Laha:2020ivk,Berteaud:2022tws,Siegert:2021upf, PhysRevD.104.023516, PhysRevLett.125.101101} have been considered using data from various satellites like {\sc Xmm-Newton} \cite{Malyshev:2022uju}. Data from this satellite have previously been used to constrain feebly interacting particles  \cite{DelaTorreLuque:2023huu,DelaTorreLuque:2023nhh} and sub-GeV DM \cite{DelaTorreLuque:2023olp, Cirelli:2023tnx, Cirelli:2020bpc} as well.

For a PBH DM explanation of the 511 keV line, previous analyses, such as Ref.~\cite{Keith:2021guq} indicated that the black hole number density of the local halo would yield a median distance that falls approximately within the confines of our Solar System. Considering the expected PBH velocity in our galactic halo, it is plausible that we could anticipate that PBHs could travel through our Solar System relatively frequently. Despite the proximity, detecting Hawking radiation emitted by a PBH of this mass would pose a considerable challenge. Conversely, by analyzing the widespread MeV-scale $\gamma$-ray emissions from the Milky Way halo, we could conclusively verify these types of scenarios using upcoming observations.

Here, we present new constraints on the PBH fraction from the Galactic diffuse X-ray emission, using {\sc Xmm-Newton} data, and leverage the recently reported longitude profile of the $511$~keV emission line. In addition, we revisit the PBH constraints from the local interstellar $e^{\pm}$ flux measurements from state-of-the-art propagation scenarios, using {\sc Voyager 1} data.
The main goal of this work consists of performing a more realistic computation of the various cosmic ray (CR) signals that would be produced by PBH evaporation when considering the full-fledged CR transport in the Milky Way, and evaluate the impact of uncertainties present in the propagation and injection setup on our limits. We utilize a fully numerical approach that does not require approximations and uses current state-of-the-art astrophysical ingredients.

The paper is organized as follows. In Sec.~\ref{sec:pbhevaporation} we cover the fundamentals of PBH evaporation, in Sec.~\ref{sec:propagation} we discuss $e^\pm$ propagation from particle injections by PBHs, in Sec.~\ref{sec:results} we discuss our main results and finally in Sec.~\ref{sec:conclusions} we conclude.

\section{PBH evaporation}
\label{sec:pbhevaporation}
Black holes are known to evaporate over time and emit particles with masses below or comparable to the temperature $T$ of the black hole through Hawking radiation~\cite{Hawking:1975vcx}. This temperature is directly related to the mass $M$ of the black hole and its spin parameter $a \equiv J/M$ with $J$ being its angular momentum~\cite{Arbey:2019jmj} (with $\hbar = c = k_B = G = 1$):
\begin{equation}
    \label{eq:BHtemp}
    T = \frac{1}{2\pi} \left(\frac{r_+ - M}{r_+^2 + a^2}\right)\;,
\end{equation}
where $r_+ \equiv M + \sqrt{M^2 - a^2}$ is the Kerr black hole horizon radius. For Schwarzschild black holes ($a = 0$), we retrieve $T = 1/(8\pi M)$. 
Then, the emission spectrum of primary particles $i$ is given by
\begin{equation}
    \frac{d^2N_i}{dtdE_i} = \frac{1}{2\pi} \sum_\text{d.o.f.} \frac{\Gamma_i(E_i,M,a^\star)}{e^{E'_i/T}\pm 1}\;,
    \label{eq:prim_spec_BH}
\end{equation}
where $a^\star \equiv a/M <1$ (if $a^\star =1$, then $T=0$, which is forbidden by thermodynamics) is the reduced spin parameter, $E'_i \equiv E_i - m\Omega$ is the energy of the emitted particle where $\Omega \equiv a^\star/(2r_+)$ is the black hole horizon rotation velocity and $m = \{-l, ..., l\}$ the projection on the black hole axis of the particle angular momentum $l$. The $\pm$ signs depend on the spin of the particles radiated: $+$ for fermions and $-$ for bosons. Finally, $\Gamma_i$ are the so-called `greybody factors' which encode the deviation from black-body physics, since emitted particles have to escape the gravitational well of the black hole. The sum is performed over the degrees of freedom (d.o.f.) of the emitted particles (spin, color, helicity and charge multiplicities). In order to compute the spectra of primary particles, we use the numerical code \verb|BlackHawk v2.2|~\cite{Arbey:2019mbc, Arbey:2021mbl}.

Then, evaporated particles can hadronise, decay or emit soft radiations. \verb|BlackHawk| also has the possibility of dealing with such processes by including tables from the particle physics codes \verb|PYTHIA|~\cite{Sjostrand:2014zea}, \verb|Herwig|~\cite{Bellm:2015jjp} and \verb|Hazma|~\cite{Coogan:2019qpu}. However, their domains of validity differ, \verb|PYTHIA| and \verb|Herwig| handle  processes with particle energies above 10 GeV very well, whereas \verb|Hazma| excels below the QCD scale ($\sim$ 250 MeV). Since we are interested in the production of sub-GeV $e^\pm$, we only use \verb|Hazma| to treat secondary processes, and limit ourselves to its domain of validity, which corresponds to a black hole mass range of $M \gtrsim 10^{14.5}$ g. An upper limit on the black hole mass can also be defined when the evaporation into $e^\pm$ is not possible anymore ($T \ll m_e$) corresponding to $M \lesssim 10^{17.5}$ g. In this black hole mass range, $e^\pm$, $\nu_{e,\mu,\tau}$ and $\gamma$ are emitted through evaporation, and to some extent (for lower black hole masses) $\mu^\pm$ and $\pi^{0,\pm}$. \verb|Hazma| handles the decay of $\mu^\pm$ and $\pi^{0,\pm}$, as well as final state radiation from all charged particles.  From now on, $\frac{d^2N_i}{dtdE_i}$ will describe the emission spectra of secondary particles.

Although it is believed that PBHs cannot acquire a substantial spin from their production process~\cite{Harada:2020pzb} unless formed in the matter-dominated universe~\cite{Harada:2017fjm}, it has been argued that they can do so by repeatedly merging with other black holes. Moreover, black holes can also acquire a spin due to the accretion of gas surrounding them. The maximum spin value a black hole can obtain through this process is $a_\textrm{lim}^\star \approx 0.998$~\cite{1974ApJ...191..507T}, known as the Thorne limit, and can slightly vary depending on the considered accretion model~\cite{Sadowski:2011ka}. A similar limit can be derived for black hole mergers~\cite{Kesden:2009ds}. Nevertheless, PBHs formed during the matter-dominated universe could evade these limits, providing a smoking gun signature of their existence.
In our study, we decide to remain agnostic on the PBH production process and consider two extreme benchmarks to quantify the impact of the choice of the spin distribution on our results. We therefore examine the case where all PBHs are Schwarzschild black holes, and another one where they are all near-extremal Kerr black holes with a spin of $a^\star = 0.9999$.

In Figs.~\ref{fig:evapspec1} and~\ref{fig:evapspec2} we show the spectra of secondary $e^\pm$ and $\gamma$ respectively for a range of black hole masses, and for spins of $a^\star = 0$ and $0.9999$. For increasing values of $a^\star$ the black hole evaporates more particles with higher energies, due to the transfer of angular momentum from the black hole to the emitted particles.

\begin{figure}[ht]
    \centering
    \includegraphics[width=0.5\textwidth]{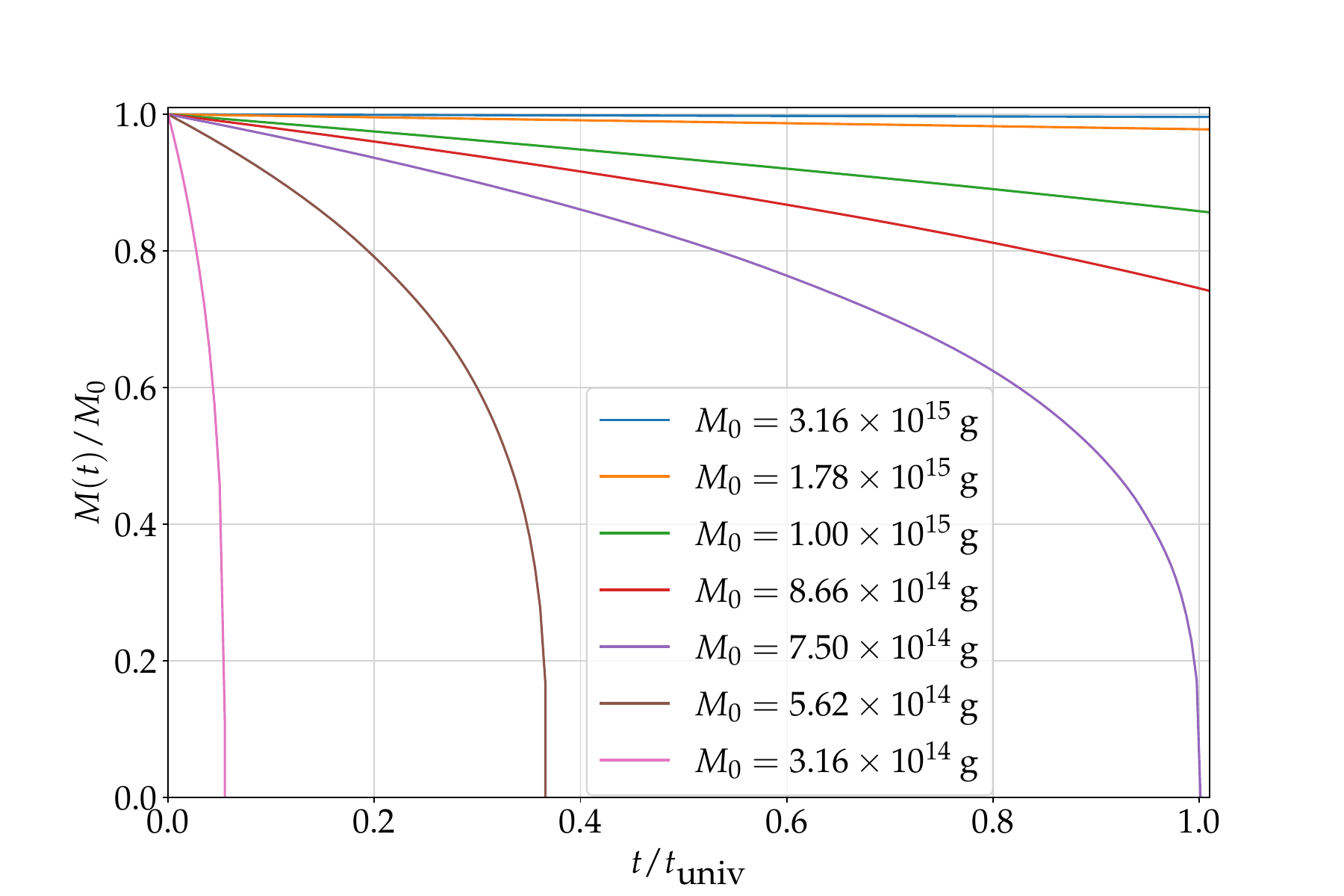}
    \caption{Evolution of the mass $M$ of Schwarzschild BHs for different initial masses $M_0$ at $t=0$. The $x$-axis represents the time in terms of fractions of the age of the Universe and $y$-axis the BHs mass in terms of fractions of its initial mass.}
    \label{fig:PBHevol}
\end{figure}

All of the discussion above only takes into account the physics from a single black hole. We then have to take into account the energy density of PBHs in our Galaxy. We investigate the possibility of PBHs constituting a fraction $f_\textrm{PBH}$ of the total amount of DM in the Universe. Therefore the number of $e^\pm$ injected by PBHs evaporation at the position vector $\vec{x}$ in our Galaxy per unit of time, energy and volume is written
\begin{equation}
    Q_e(E_e,\vec{x}) = f_\textrm{PBH}\rho_\textrm{DM}(\vec{x}) \int_{M_\textrm{min}}^\infty \frac{dM}{M}\frac{dN_\textrm{PBH}}{dM} \frac{d^2N_e}{dtdE_e}\;,
\end{equation}
where $\rho_\textrm{DM}$ is the DM energy density profile and $\frac{dN_\textrm{PBH}}{dM}$ is the mass distribution of PBHs normalized to 1, since $f_\textrm{PBH}\,\rho_\textrm{DM}$ already represents the spatial distribution of the PBH energy density in the Galaxy. $M_\textrm{min} \approx 7.5 \times 10^{14}$ g corresponds to the minimal mass of PBHs today. As shown in Fig.~\ref{fig:PBHevol}, PBHs formed in the early Universe with a mass below $M_\textrm{min}$ should have all evaporated by now, whereas PBHs with an initial mass of $10^{15}$ g have experienced their mass decreasing by $\mathcal{O}(20\%)$. Thus, we follow the approximation where all PBHs with $M_\textrm{PBH} \leq M_\textrm{min}$ do not exist today, and the remaining ones have the same mass distribution as in their formation in the early Universe.

We consider the following PBHs mass distribution
\begin{equation}
    \label{eq:lognorm}
    \frac{dN_\textrm{PBH}}{dM} = \frac{1}{\sqrt{2\pi}\sigma M}\exp\left(-\frac{\log^2(M/M_\textrm{PBH})}{2\sigma^2}\right)\;,
\end{equation}
where $M_\textrm{PBH}$ is the peak PBH mass, and $\sigma$ is the standard deviation of the distribution. This mass function is relevant when assuming the formation of PBHs from a large enhancement in the inflationary power spectrum~\cite{Dolgov:1992pu,Carr:2017jsz,Balaji:2022rsy,Qin:2023lgo}. We explore values of $\sigma$ varying from 0 to 2, the case $\sigma \rightarrow 0$ corresponding to a monochromatic mass distribution ($\frac{dN_\textrm{PBH}}{dM}=\delta(M-M_\textrm{PBH})$).

\section{Electron-positron propagation and diffuse emissions}
\label{sec:propagation}

\subsection{Electron-positron propagation}

Light asteroid-mass PBHs evaporating in the Galaxy produce a continuous injection of electrons and positrons that eventually leads to a diffuse and steady-state flux peaking at energies above the MeV scale. Once injected, as described in Eq.~\eqref{eq:prim_spec_BH}, these particles propagate and interact with the Galactic gas, magnetic fields and ambient light. For particles propagating with energies below a few tens of MeV the main process is energy losses, dominantly from ionisation of neutral gas, although bremsstrahlung and Coulomb interactions with the ambient plasma are relevant too~\cite{DelaTorreLuque:2023olp, Boudaud:2016mos}. Diffuse reacceleration~\cite{osborne1987cosmic, seo1994stochastic}, that is the product of energy exchange between relativistic charged particles and the perturbations of the plasma, also becomes very relevant at such energies, since the low mass and energy of the $e^\pm$ produced from these PBHs can be easily boosted to much higher energies, even reaching GeV energies~\cite{DelaTorreLuque:2023olp}.
All of the mentioned processes are encoded in the diffusion-advection-convection-loss equation~\cite{Ginz&Syr, DRAGON2-1}
\begin{widetext}
    \begin{equation}
    \label{eq:CRtransport}
        - \nabla\cdot\left(D\vec{\nabla} f_e + \vec{v}_c f_e \right) - \frac{\partial}{\partial p_e} \left[\dot{p}_e f_e - p_e^2 D_{pp} \frac{\partial}{\partial p_e}\left(\frac{f_e}{p_e^2}\right) - \frac{p_e}{3}\left(\vec{\nabla}\cdot\vec{v}_c f_e\right)\right] = Q_e\;,  
    \end{equation}
\end{widetext}
which has to be solved for $f_e \equiv \frac{dn_e}{dp_e}$, the density of $e^\pm$ per unit momentum {at a given position. This equation includes: i) spatial diffusion with coefficient $D$, ii) momentum losses $\dot{p}_i$ due to interactions with the galactic environment, iii) momentum diffusion (or reacceleration) with diffusion coefficient $D_{pp}$, iv) convection due to the galactic wind velocity $\vec{v}_c$. We solve Eq.~\eqref{eq:CRtransport} using the numerical code \verb|DRAGON2|~\cite{DRAGON2-1,DRAGON2-2}. As a benchmark, we use a Navarro-Frenk-White (NFW)~\cite{Navarro:1995iw} DM distribution. 
We have computed tables with the $e^{\pm}$ injection spectra from PBH evaporation ($\frac{d^2N_e}{dtdE_e}$) using \verb|BlackHawk|. These are an input in the \verb|DRAGON2| code and are used analogously to the case of decaying DM. These tables are available upon request from the authors and will be released along with a new public version of \verb|DRAGON2|.

In our benchmark setup, the diffusion parameters are the same as prescribed in Ref.~\cite{DelaTorreLuque:2023olp}, where we refer the reader for full details. These parameters are obtained from detailed fits of existing CR data at Earth. However, given that systematic uncertainties are difficult to asses and different CR analyses can find slightly different results~\cite{Strong_1998, osborne1987cosmic, seo1994stochastic}, we consider other extreme propagation scenarios that allow us to evaluate the impact of these uncertainties in the predicted $e^{\pm}$ spectra and the bounds on the fraction of DM constituted by PBHs.
\begin{figure*}[th]
    \centering
    \includegraphics[width=0.49\linewidth]{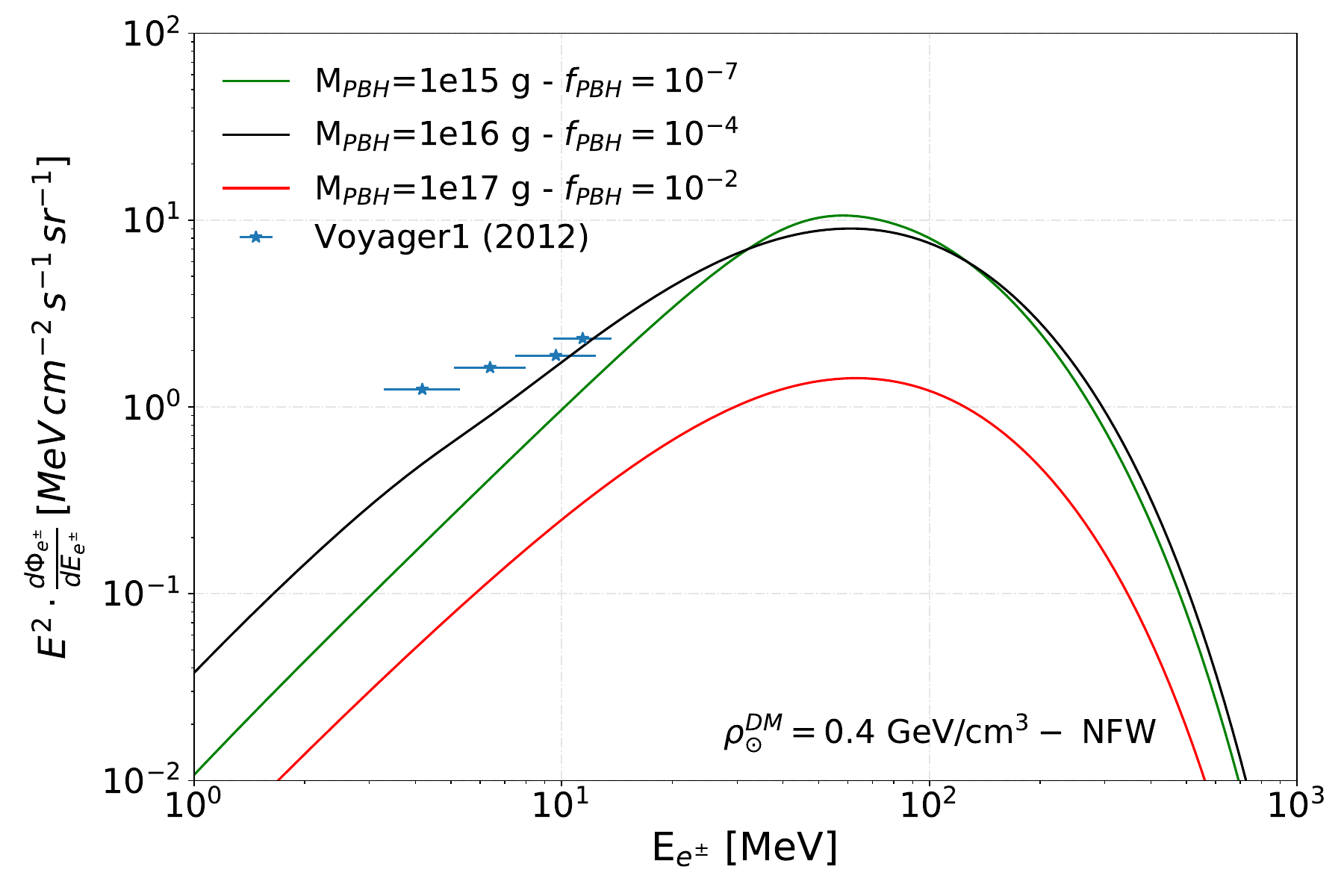}
    \includegraphics[width=0.49\linewidth]{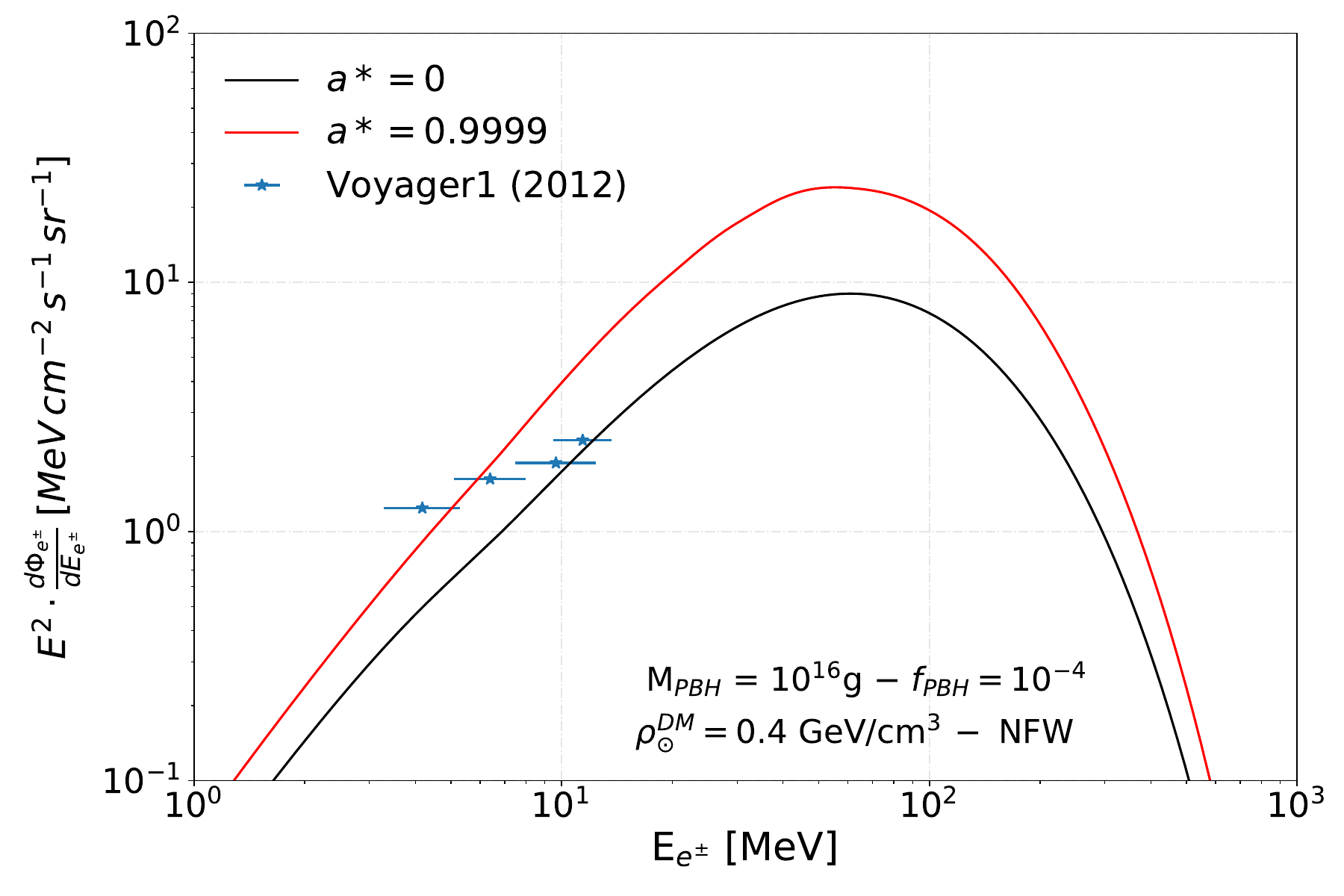}
    
    \includegraphics[width=0.49\linewidth]{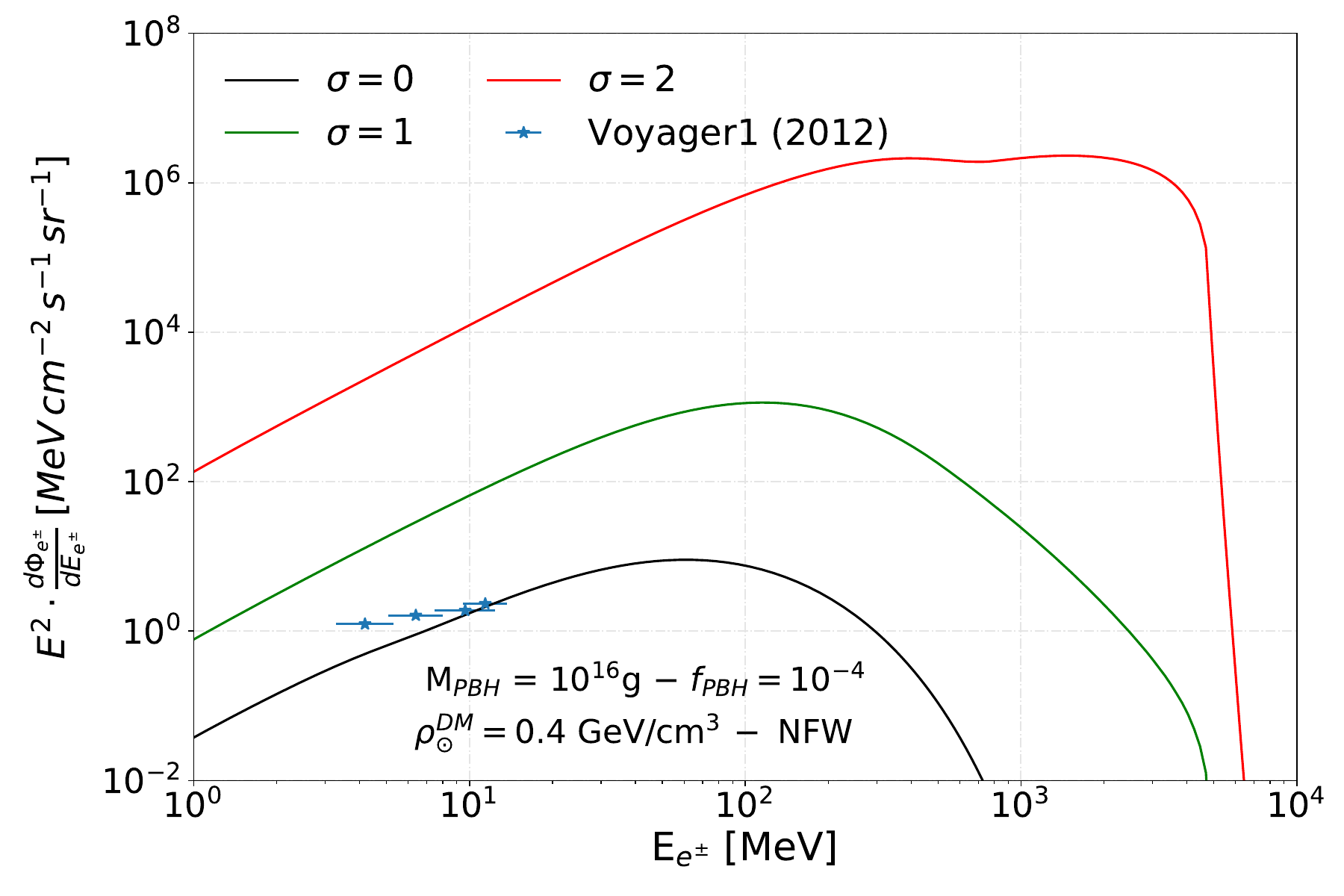}
    \includegraphics[width=0.49\linewidth]{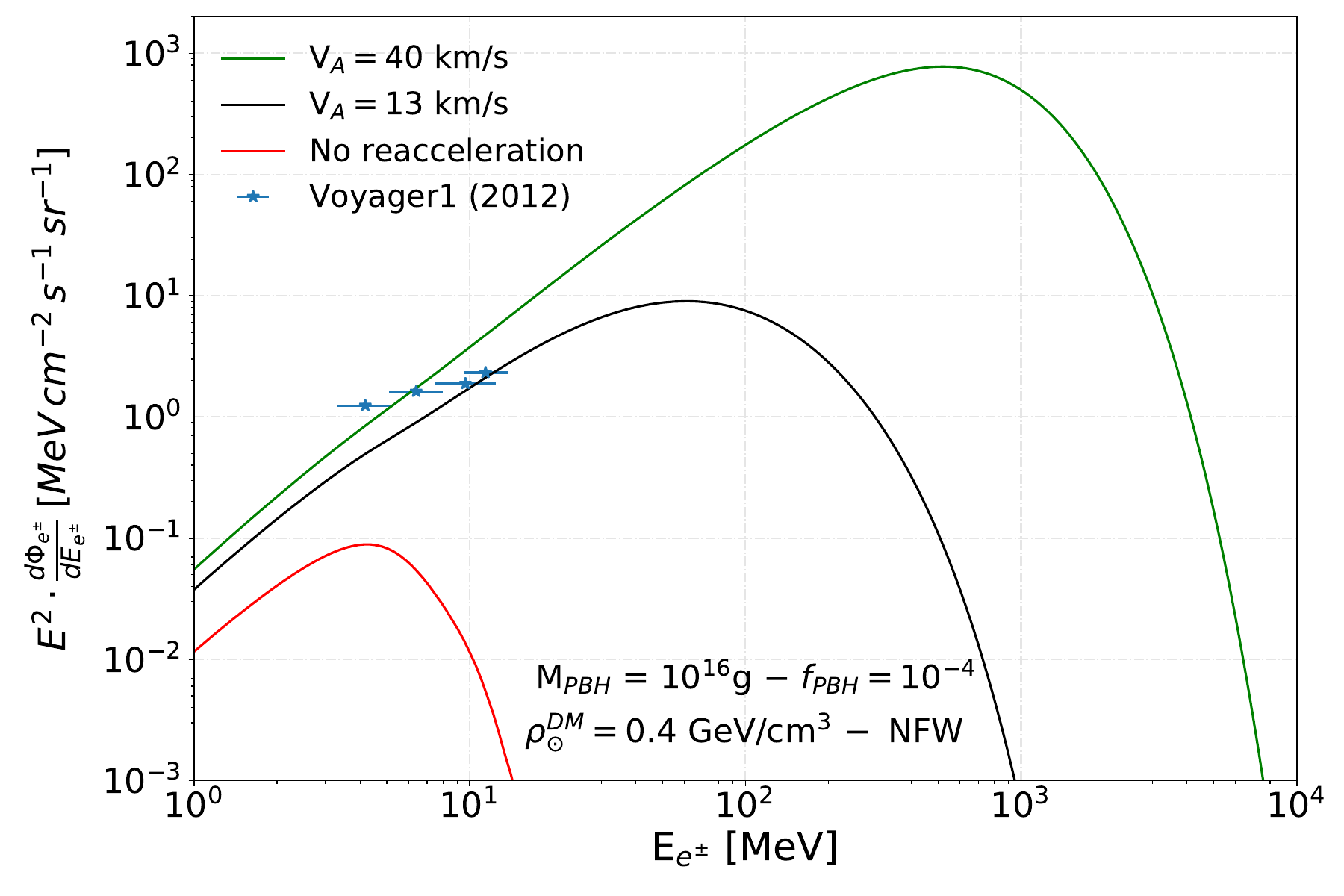}
    \caption{Local $e^{\pm}$ spectrum ($e^+$ + $e^-$) generated from the evaporation of light PBHs under different assumptions. \textbf{Top-left:} Comparison of the spectrum for Schwarzschild PBH of different masses, assuming their mass distribution to be monochromatic. \textbf{Top-right:} Comparison of the expected local $e^{\pm}$ spectrum from Schwarzschild ($a^\star=0$) and near-extremal Kerr ($a^\star=0.9999$) PBHs, assuming they are monochromatic in mass. \textbf{Bottom-left:} Comparison of the predicted local $e^{\pm}$ spectrum for Schwarzschild PBHs assuming a log-normal mass distribution with different values of the standard deviation $\sigma$. \textbf{Bottom-right:} Comparison of the local $e^{\pm}$ spectrum for monochromatic Schwarzschild PBHs, for different levels of reacceleration (i.e. values of $V_A$).}
    \label{fig:VoyagervsVa}
\end{figure*}
We first repeat our calculations for `realistic' variations of the propagation parameters found in our analysis, which consist of taking the values that maximize the difference in flux from the benchmark case at $3\sigma$. Similar to the case of DM decay, the parameters with a greater effect on the diffuse spectra produced from PBHs are the Alfvén velocity $V_A$ parameter that controls the level of diffuse reacceleration~\cite{DelaTorreLuque:2023nhh, DelaTorreLuque:2023olp} and the height $H$ of the halo, which dictates the volume where CRs are confined and where PBHs produce particles that can reach us. In this way, to obtain a realistic uncertainty band in our predictions, we use a conservative setup where $H = 4$~kpc and $V_A=7$~km/s, which produces a lower (and therefore more conservative) flux of $e^\pm$. In turn, the more aggressive setup is meant to increase the flux of $e^\pm$ from PBHs, and uses values of $H = 12$~kpc and $V_A=20$~km/s. As a point of reference, we recall that our benchmark values are $H = 8$~kpc and $V_A=13.4$~km/s. 
We tested an even more `general' and extreme variation of propagation setup, that ensures that the flux of particles must be between the two extremes: The `optimistic' case, where $H = 16$~kpc and $V_A=40$~km/s, is much higher than the typical values\footnote{We consider that $V_A=40$~km/s is the maximum realistic value for $V_A$, since it already implies that most of the injected energy of CRs are coming from the perturbations of the interstellar plasma and not from supernova remnants, which will break the standard paradigm of CR propagation, see Ref.~\cite{Drury:2016ubm, drury2015cosmicray}}. Then, the most `pessimistic' case will be that with no reacceleration ($V_A=0$~km/s) and $H = 3$~kpc. These two cases are unlikely, given the fact that propagating CRs implies energy exchange with plasma waves and therefore non-zero reacceleration, and values below $H = 3$~kpc seem to be strongly disfavoured from CR analyses and other existing constraints~\cite{DeLaTorreLuque:2021yfq, Evoli_2020, Weinrich_H, delaTorreLuque:2022vhm}.
As an example, we show in the bottom-right panel of Fig.~\ref{fig:VoyagervsVa} the dramatic effect of reacceleration in the $e^\pm$ spectrum at Earth from a $M_\textrm{PBH}=10^{16}$~g PBH, for different values of $V_A$, comparing our benchmark scenario with the aforementioned optimistic and pessimistic ones.

\begin{figure*}[ht!]
    \centering
    \includegraphics[width=0.49\linewidth]{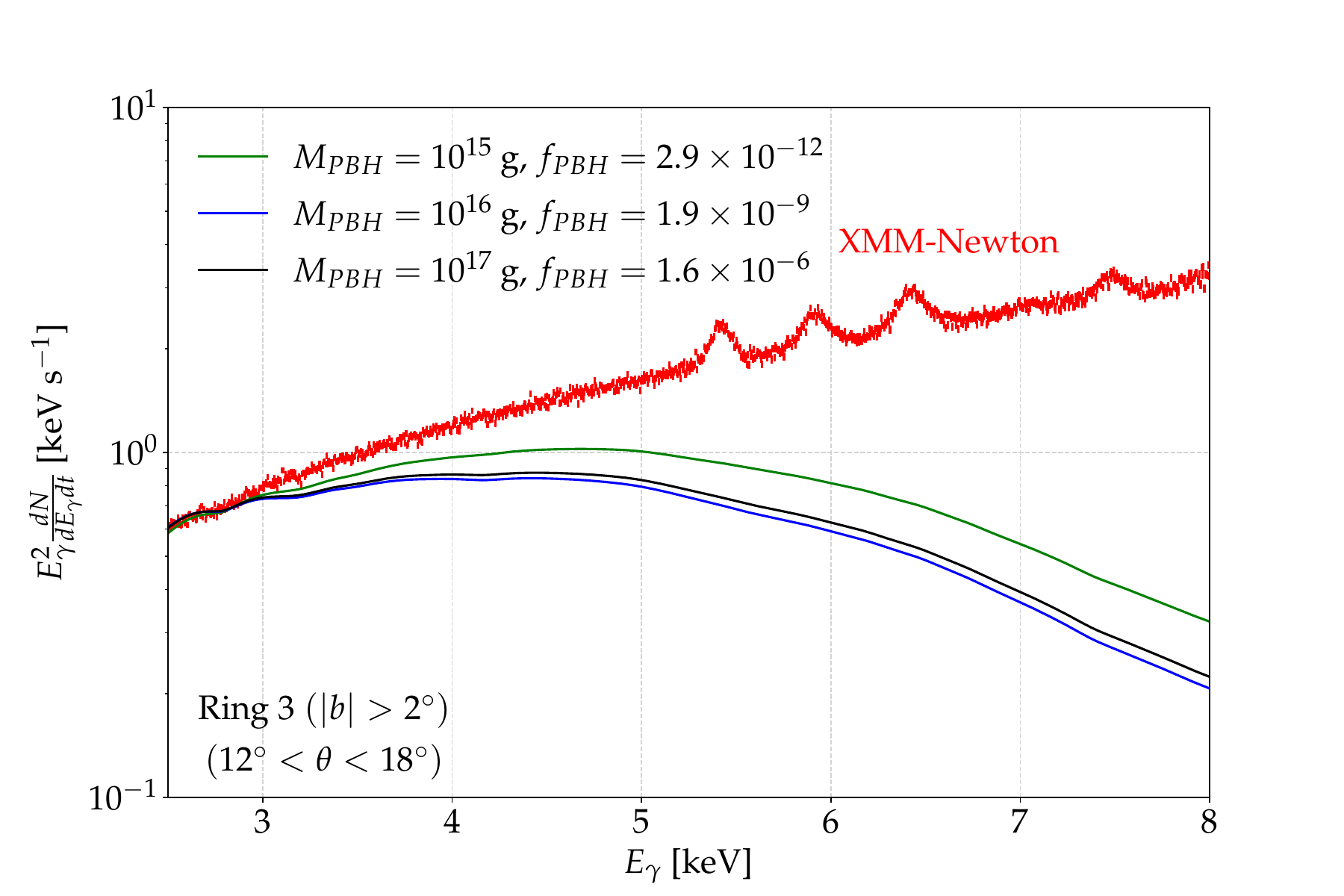}
    \includegraphics[width=0.49\linewidth]{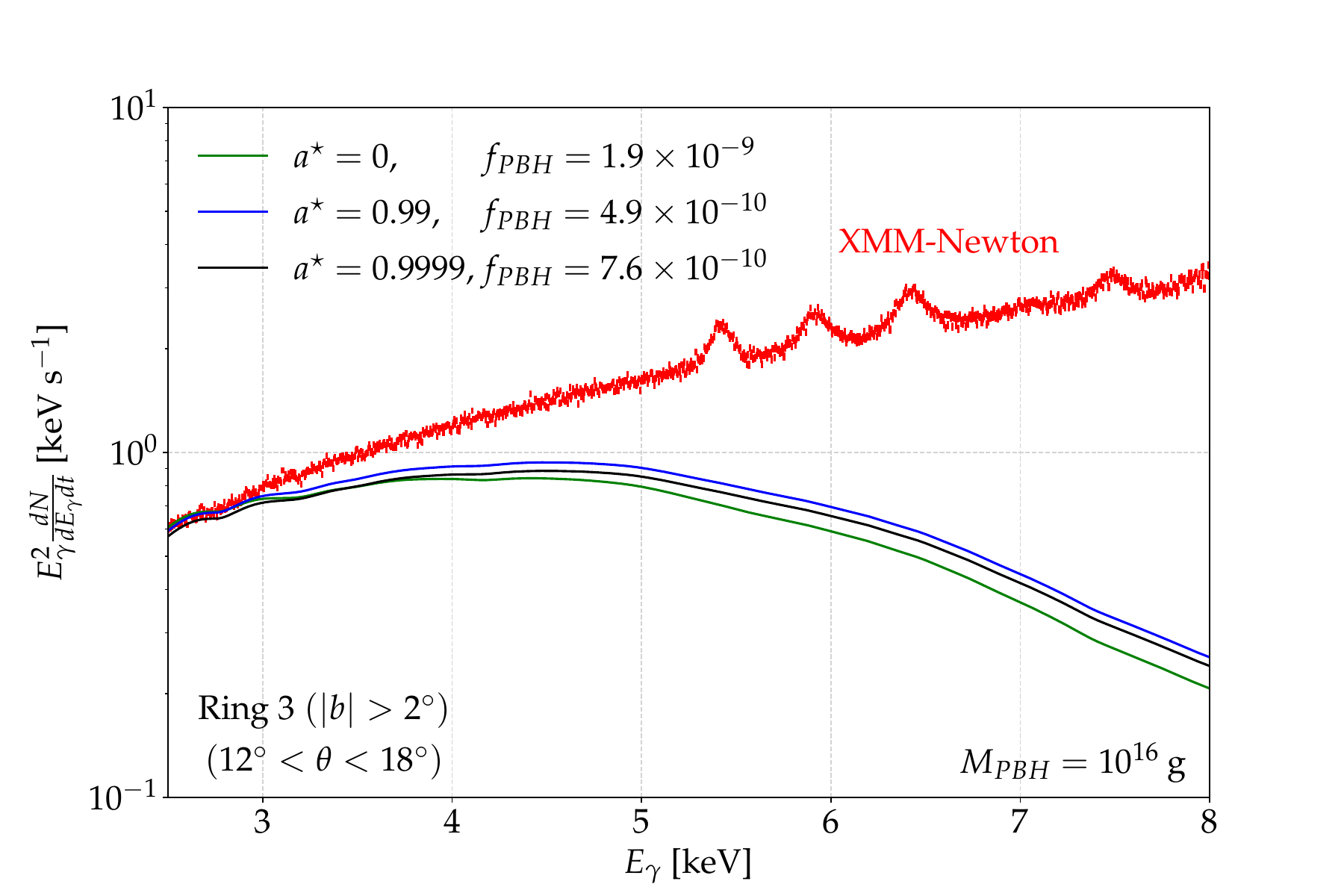}
    \includegraphics[width=0.49\linewidth]{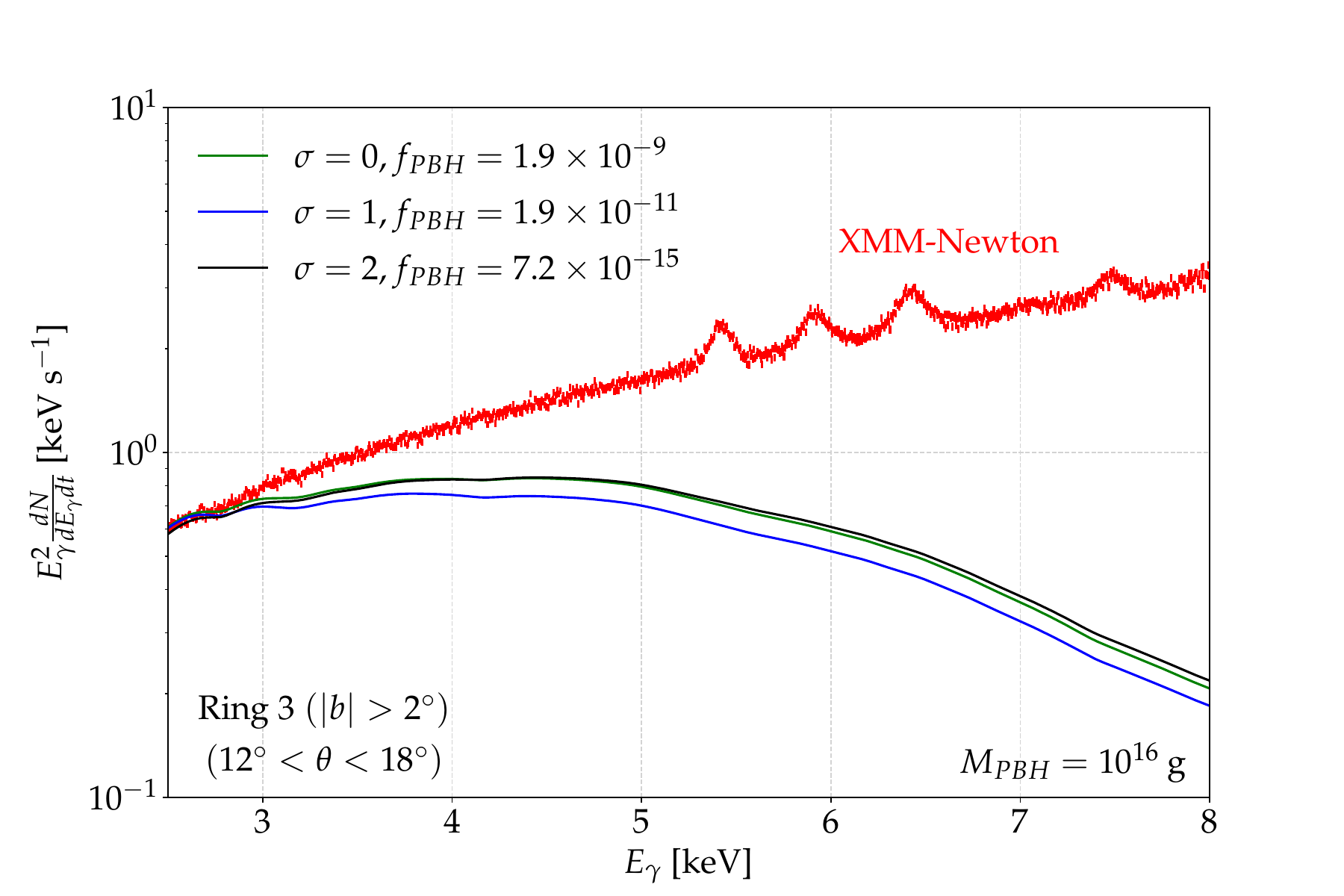}
    \caption{Comparison of the predicted DM-induced X-ray emission with diffuse X-ray data from {\sc Xmm-Newton} in a region close to the Galactic Center. We show the prediction is shown for different values of $M_\textrm{PBH}$ when the PBH mass distribution is monochromatic (upper left panel), of $a^\star$ (upper right panel) and $\sigma$ when the distribution is log-normal (lower panel).}
    \label{fig:XMMComp}
\end{figure*}

We will obtain constraints on the fraction of DM that can be in the form of PBHs using {\sc Voyager 1}~\cite{cummings2016galactic, stone2013voyager} measurements of the local  flux. A comparison of the total electron-positron (i.e. $e^+$ + $e^-$) flux measured by {\sc Voyager 1} and the predicted local electron spectrum for monochromatic PBHs of different masses, for our benchmark propagation setup and NFW DM profile, is shown in the top-left panel of Fig.~\ref{fig:VoyagervsVa}.
In addition, we also illustrate the spectra predicted assuming a log-normal PBH mass distribution (see Eq.~\eqref{eq:lognorm}) with $\sigma=1$ (green line) and $\sigma=2$ (red line) in the bottom-left panel. This allows one to see how the $\sigma$ parameter affects our predictions, given that physically one must expect a non-zero $\sigma$. It can be seen that the higher $\sigma$ is, the higher is the expected  flux and the higher is the energy reached by the electrons. The main reason for this is that the contribution from lower-mass PBHs is very important and dominates the spectra of these particles. 

Then, in the top-right panel we compare the spectra produced from PBHs with different values of the spin parameter $a^\star$. In particular, we show the cases of Schwarzschild ($a^\star=0$) and near-extremal Kerr ($a^\star=0.9999$) PBHs, as well as for $a^\star=0.99$. As one can see from the figure, spin of the PBH always leads to a higher flux produced by PBH evaporation although not changing its spectral shape appreciably, in agreement with what was found in Refs.~\cite{arbey2020primordial, Arbey:2021mbl}.

\subsection{Diffuse X-ray emission}

During their propagation, the population of electron-positron pairs injected in the Galaxy produce different secondary radiations that can be used to track their distribution and density.  Especially, their interaction with the Galactic magnetic field will generate a diffuse synchrotron emission that can be observed at kHz-MHz. On top of that, they will interact via bremsstrahlung with the ionized gas in the interstellar medium (ISM), leading to $\gamma$-ray radiation at the MeV scale. Here, we focus on the X-ray diffuse emission produced from the inverse Compton interaction of this  population with the Galactic radiation fields (mainly the cosmic microwave background, optical and UV light from stars and infrared from the scattering of the latter on dust), because of the high constraining power of X-ray measurements from {\sc Xmm-Newton}, as mentioned above.

\begin{figure*}[ht!]
    \centering
        \includegraphics[width=0.49\linewidth]{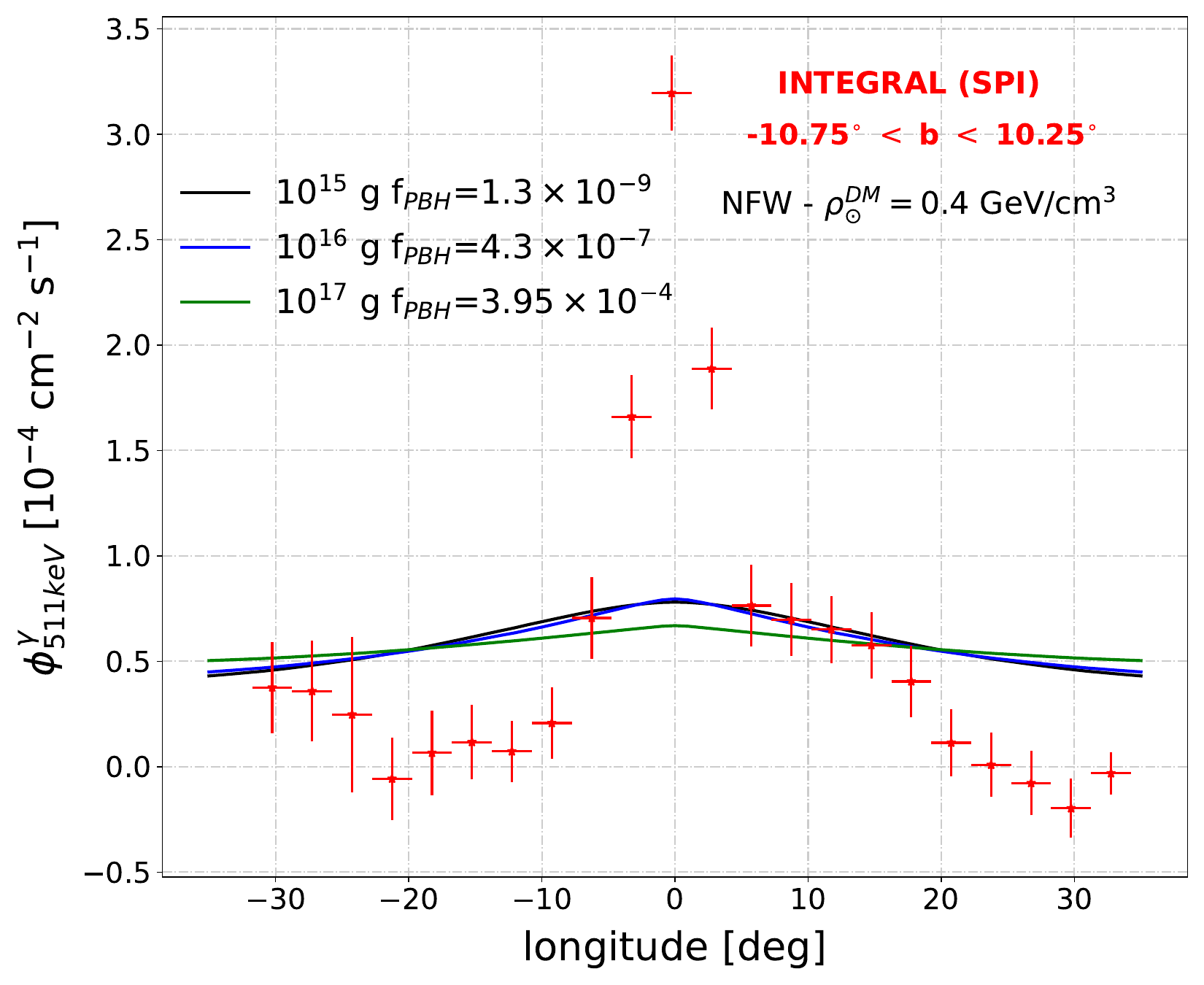}
    \includegraphics[width=0.49\linewidth]{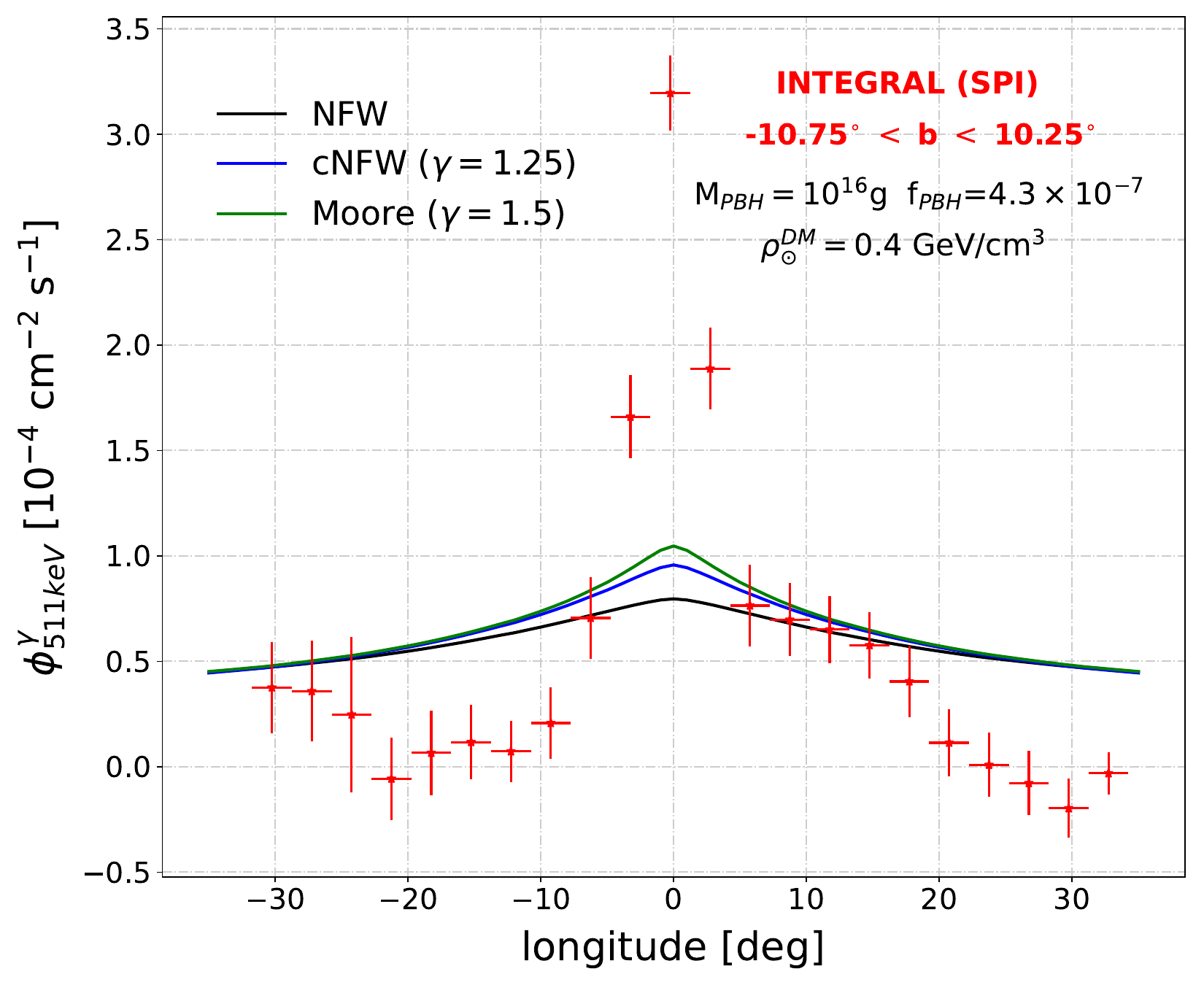}
    \caption{Comparison of the expected longitude profile of the $511$~keV emission from PBH evaporation with {\sc Integral/Spi} data. The left panel shows a comparison of the predicted profile for different DM distributions, for the case of Schwarzschild PBH monochromatically distributed with $M_\text{PBH}=10^{16}$~g, while the right panel shows a comparison of the signals expected from different PBH masses, assuming again Schwarzschild and monochromatically distributed PBHs.}
    \label{fig:511_profs}
\end{figure*}

To calculate the diffuse X-ray emission generated from the diffuse (steady-state) distribution of $e^\pm$ in the Galaxy that we obtain with \verb|DRAGON2|, we employ the \verb|Hermes| code~\cite{Dundovic:2021ryb}, a software designed to compute full maps of nonthermal radiative processes such as radio, X-ray and $\gamma$-rays, as well as neutrino emissions. The total X-ray flux also includes photons directly emitted during PBH evaporation, as well as final state radiations produced by evaporated $e^\pm$, $\mu^\pm$ and $\pi^\pm$. It turns out that this component is sub-dominant compared to the X-ray flux emitted during the transport of evaporated $e^\pm$.
We compute $2\sigma$ bounds from the diffuse Galactic X-ray emission observed by {\sc Xmm-Newton}~\cite{XMMwebpage, XMM, Balaji:2025afr} in the $2.5-8$~keV band, as done in Ref.~\cite{DelaTorreLuque:2023olp, Cirelli:2023tnx}, where we refer the reader for more details. In Fig.~\ref{fig:XMMComp}, we compare the X-ray diffuse emission expected from PBH evaporation. In the top-left panel, we show the case of monochromatic PBHs with masses of $10^{15}$~g, $10^{16}$~g and $10^{17}$~g, for our benchmark propagation setup and NFW DM profile.
In the top-right panel, we compare the emission expected from a Schwarzschild PBH ($a^\star=0$) and Kerr PBHs with $a^\star=0.99$ and the extreme case of $a^\star=0.9999$, all for $M_\textrm{PBH}=10^{16}$ g. Note that the fraction of PBHs comprising DM is different for every case, as indicated in the legend.
In the bottom panel, we show results for Schwarzschild PBHs distributed log-normally with a mean mass of $10^{16}$~g and different values of $\sigma$, ranging from $\sigma=0$ (monochromatic case) to the wider $\sigma=2$. The conclusions for the impact of these parameters in the X-ray Galactic diffuse emission are similar to those found for the local flux of Fig.~\ref{fig:VoyagervsVa}. We finally remark that the associated bremsstrahlung emission is negligible at keV energies.


\subsection{511 keV line profile}

As the injected positrons are propagating in the Galaxy, they lose their energy until they eventually reach thermal energies of the medium that they are travelling through. After a typical time scale of $0.1-10$~Myr~\cite{Guessoum:2005cb, Guessoum1991ApJ}, thermal positrons will form a positronium state with ambient electrons and decay into a pair of $511$~keV photons with $25\%$ probability (through the para-positronium state), that lead to a bright line emission. Assuming that the thermal positrons follow the distribution of the steady-state diffuse positrons injected by PBHs, we have calculated the intensity and distribution of the $511$~keV line from low mass DM annihilating and decaying in Ref.~\cite{DelaTorreLuque:2023cef} and this case leads to very similar signals to those expected from PBH evaporation.

In this work, we obtain constraints from the longitude profile of the $511$~keV line emission following the procedure described in Ref.~\cite{DelaTorreLuque:2024zsr}, where the profile of the line is directly proportional to the distribution of propagated (steady-state) positrons. 
We have tested that our results are compatible with previous evaluations applied to other exotic sources of positrons~\cite{Calore:2021klc, Calore:2021lih, DelaTorreLuque:2023huu, DelaTorreLuque:2023nhh, Carenza:2023old}.

\begin{figure*}[ht]
    \centering
    \includegraphics[width=0.65\linewidth]{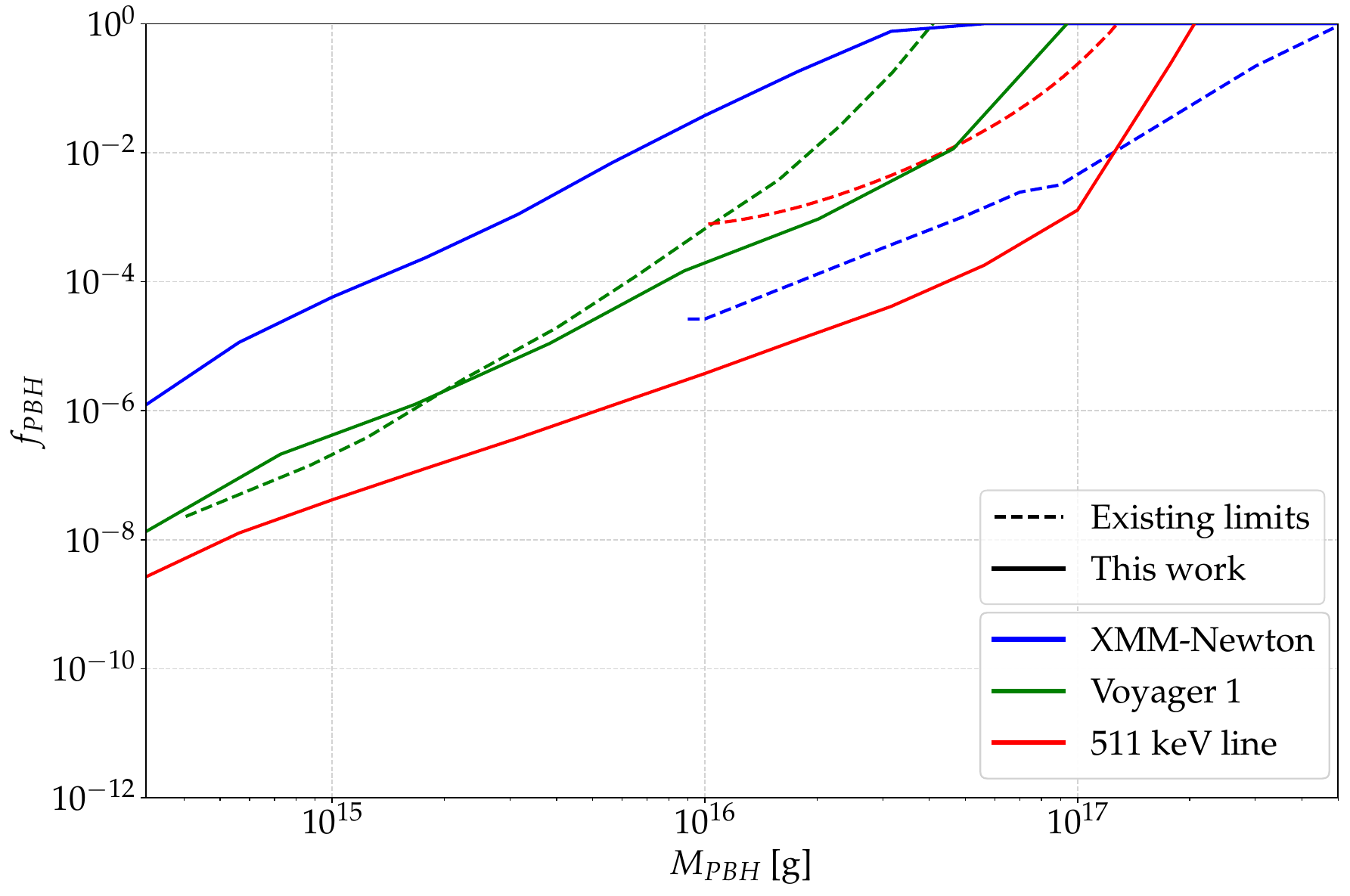}
    \cprotect\caption{Comparison of the 95\% confidence limits on $f_\textrm{PBH}$ derived in this work with other existing ones. The color of the lines represent the different probes used to set the constraints: green for the $e^\pm$ measurements from {\sc Voyager 1}, blue for X-ray diffuse observations, red for the 511 keV excess reported by {\sc Integral}. The two different line styles correspond to either the bounds derived either in this work (solid) or in the literature~\cite{Boudaud:2018hqb,Laha:2019ssq,Tan:2024nbx,Mittal:2021egv} (dashed) reported in \verb|PBHbounds|~\cite{kavanagh_2019}.}
    \label{fig:moneyplot}
\end{figure*}

In the left panel of Fig.~\ref{fig:511_profs} we show the predicted longitude profile of the $511$~keV emission for PBHs of masses between $10^{15}$ and $10^{17}$~g compared to {\sc Integral/Spi} data~\cite{Siegert:2015knp}, assuming a NFW profile and with the PBH fraction of DM $f_\textrm{PBH}$ specified in the legend for each case. It can be seen that the most constraining data points are those obtained at high longitudes. Given that these points are also those expected to be more affected by systematic uncertainties (mainly background noise and the need of templates to extract measurements) and the limited statistics of the measurements, the bounds that we derive are conservatively calculated 
 by multiplying by a factor of $2$, as a proxy for the effect of systematic uncertainties, as done in Refs.~\cite{DelaTorreLuque:2023nhh, DelaTorreLuque:2024zsr}.
We show in the right panel of Fig.~\ref{fig:511_profs} a comparison of the predicted line profile with the NFW DM distribution with other popular DM profiles, namely a Moore profile ($\gamma=1.5$)~\cite{Moore_1999}, a contracted NFW profile similar to the one fitting the Galactic Center excess ($\gamma=1.25$)~\cite{Ackermann_2017}, for a monochromatic $10^{16}$~g mass PBH. A Burkert~\cite{Burkert:1995yz}, or other cored DM distributions~\cite{Cirelli:2010xx}, will simply lead to a flatter profile. As one can see, the predicted profiles are very similar at high longitudes and only change significantly around the central longitudes. Therefore, the uncertainties in the derived constraints from the DM distribution are very small.
Similarly, a different spin or adoption of $\sigma\neq0$ has no significant consequences on the shape of the profile and essentially changes only the  normalization of the signal.

\section{Results and comparison with other work}
\label{sec:results}

In this section, we discuss the limits on PBHs we derived in this work, for our benchmark scenario. In addition to displaying the limits using the three probes ($e^\pm$, 511 keV line and diffuse $X$-rays), we also show the impact on these limits when assuming different PBH mass distributions, spin distributions and propagation models.
Here, we set $2\sigma$ bounds on $f_\textrm{PBH}$ and $M_\textrm{PBH}$ by applying the criterion 
\begin{equation}
    \sum_i \left( \frac{\textrm{Max}\left[\phi_{\textrm{PBH}, i}(M_\textrm{PBH}) - \phi_{i}, 0 \right]}{\sigma_i}\right)^2 =4 \,, 
\end{equation} 
where $i$ denotes the data point, $\phi_\textrm{PBH}$ is the predicted flux induced by PBH evaporation, $\phi_{i}$ is the measured flux and $\sigma_i$ the associated standard deviation of the measurements.

In Fig.~\ref{fig:moneyplot} we show our benchmark limits on Schwarzschild PBHs, assuming a monochromatic mass distribution and a NFW DM profile, and compare them to existing ones. The solid lines represent the bounds derived in this work, while the dot-dashed lines represent some of the most stringent limits on $f_\textrm{PBH}$ reported in \verb|PBHbounds|~\cite{kavanagh_2019} across the $10^{15}-5\times10^{17}$ g PBH mass range.

\begin{figure*}[ht!]
    \centering
    \includegraphics[width=0.49\linewidth]{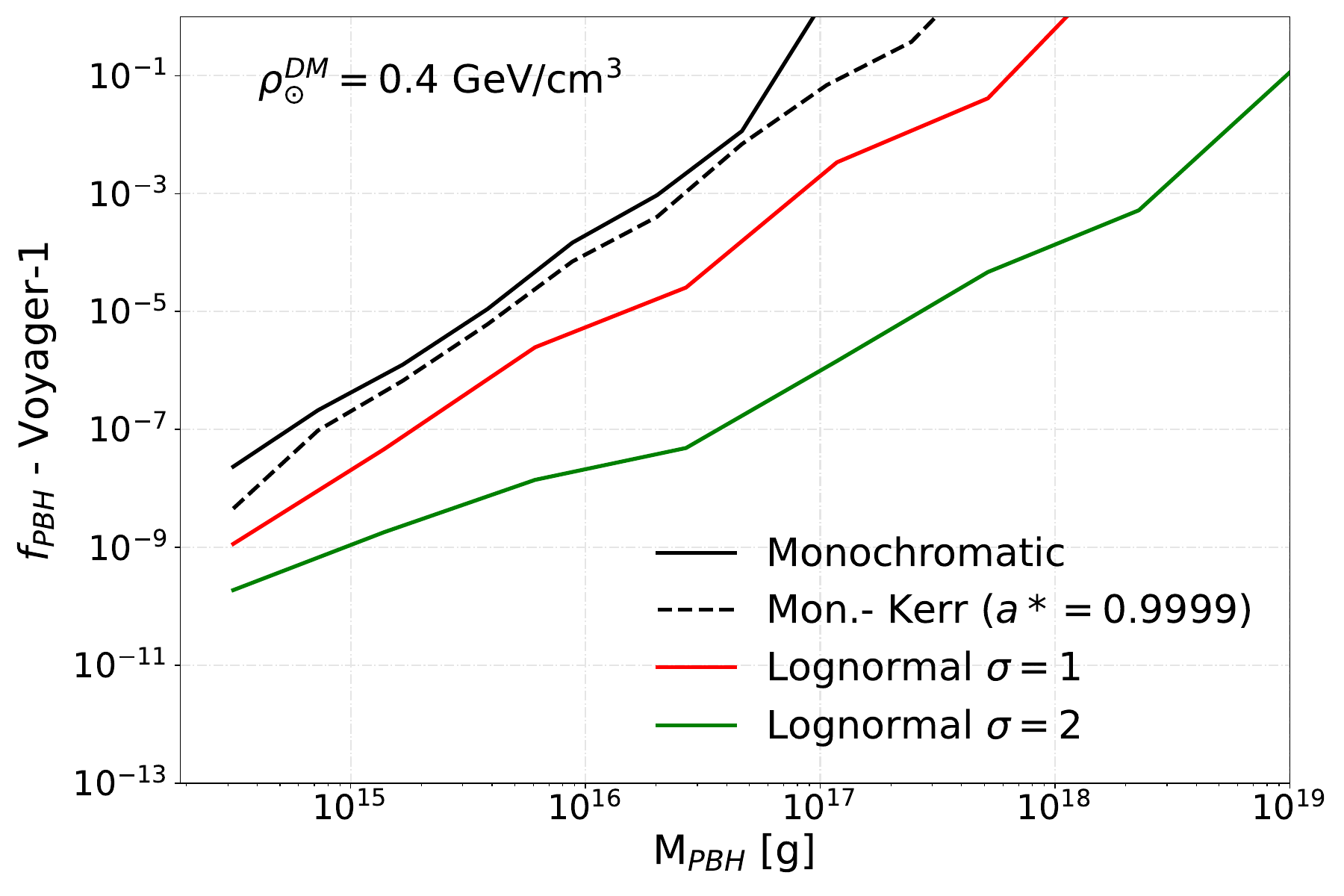}
    \includegraphics[width=0.49\linewidth]{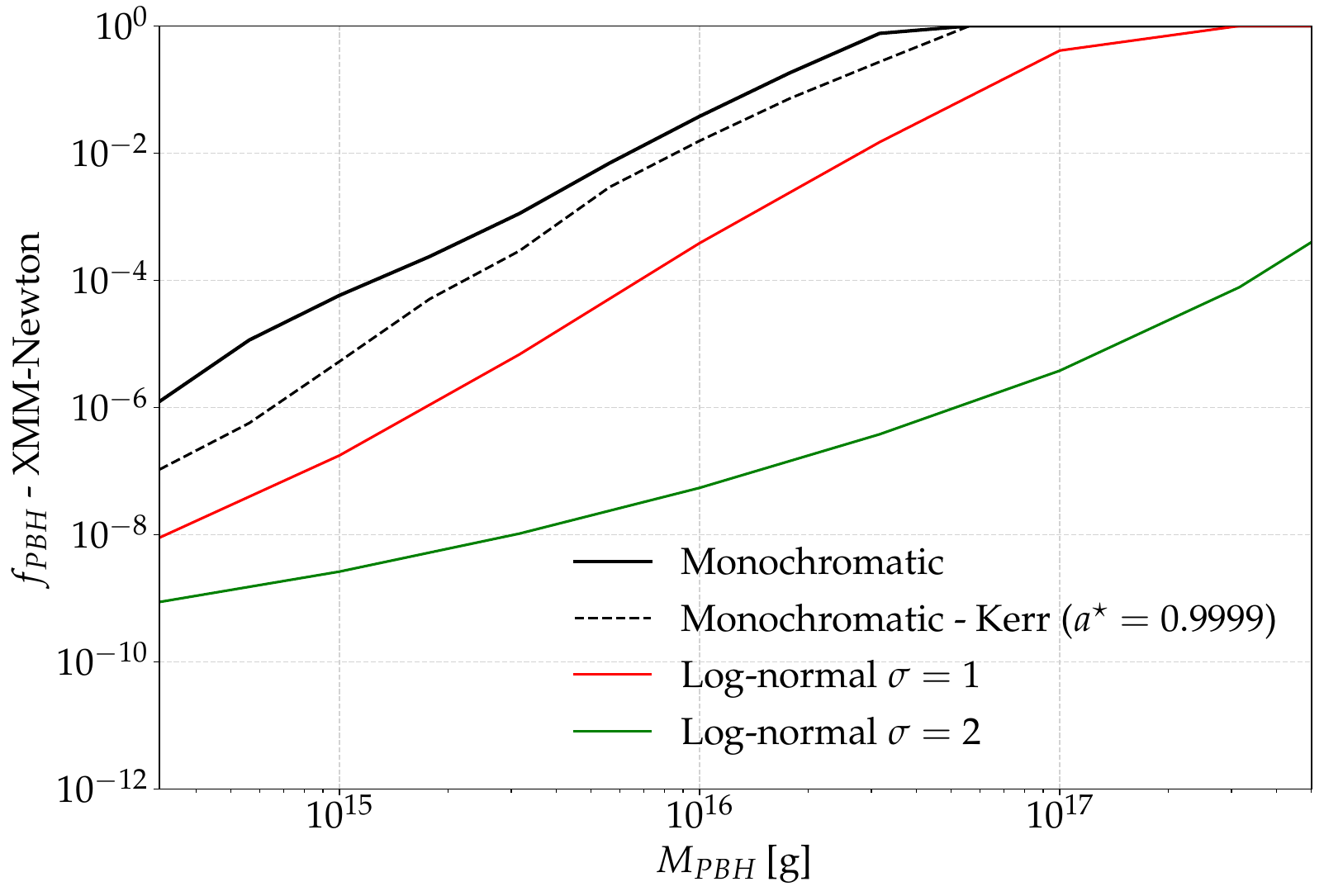}
    \includegraphics[width=0.49\linewidth]{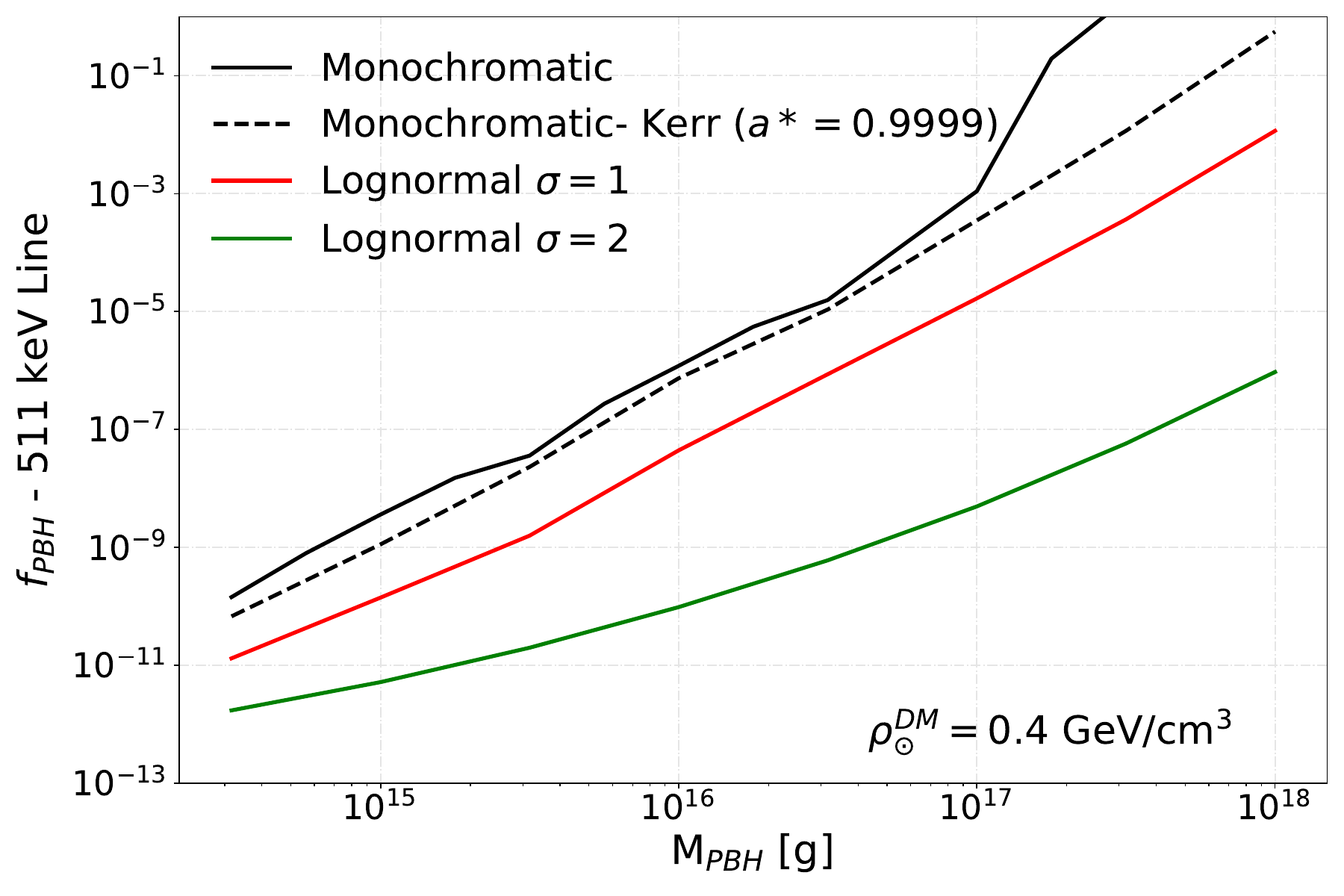}
    
    \caption{Limits we derive using {\sc Voyager 1} (upper left panel), {\sc Xmm-Newton} (upper right panel) and the 511 keV line reported by {\sc Integral} (lower panel) for mass distributions with $\sigma = 0$ (monochromatic, solid black line), 1 (red line) and 2 (green line), as well as for the extreme assumption where all PBHs have a spin of $a^\star = 0.9999$ (dashed line).}
    \label{fig:SomeLimits}
\end{figure*}

We show the {\sc Voyager 1} limits in green on Fig.~\ref{fig:moneyplot}, where the dashed line corresponds to the limit reported in Ref.~\cite{Boudaud:2018hqb} without background subtraction. The authors used a propagation model with strong reacceleration named `model B'. Our {\sc Voyager 1} limit is comparable to the existing one for $M_\textrm{PBH} \lesssim 10^{16}$ g and gets more stringent for higher PBH masses. The reason is likely due to the differences in how reacceleration is implemented in the \verb|DRAGON2| code with respect to the the semi-analytical code \verb|USINE|~\cite{Maurin_2020}\footnote{\url{https://dmaurin.gitlab.io/USINE/}} used in their work, where reacceleration only takes place in a thin disk, instead of adopting uniform reacceleration across the whole Galaxy, that is important given that CR particles spend most of their time in the Galactic halo while propagating. In addition, to model energy losses, which are key for MeV particles, \verb|USINE| needs to make use of the pinching method~\cite{Boudaud:2016jvj}.

The limits from diffuse X-ray emissions are shown in blue on Fig.~\ref{fig:moneyplot}, where the dashed line is the limit set in Ref.~\cite{Tan:2024nbx}. The authors have computed the flux of (prompt) X-ray emissions from the evaporation of extragalactic PBHs and compared it to the isotropic cosmic X-ray background measurements, without considering the secondary inverse Compton emission, to set a limit on $f_\textrm{PBH}$. 
Remarkably, the low energy part of the X-ray measurements are those most constraining. Therefore, X-ray diffuse measurements at lower energies are expected to improve these limits significantly. However, X-ray emission starts to be severely absorbed by interstellar gas, which can make it more difficult to improve these constraints using lower energy data. 
Current work in progress with {\sc eRosita} Galactic diffuse data indicate that our limit can improve by up to an order of magnitude.
Furthermore, we note that the most constraining X-ray data corresponds to the inner regions (see Refs.~\cite{DelaTorreLuque:2023olp, Cirelli:2023tnx}).

\begin{figure*}[ht!]
    \centering
    \includegraphics[width=0.465\linewidth]{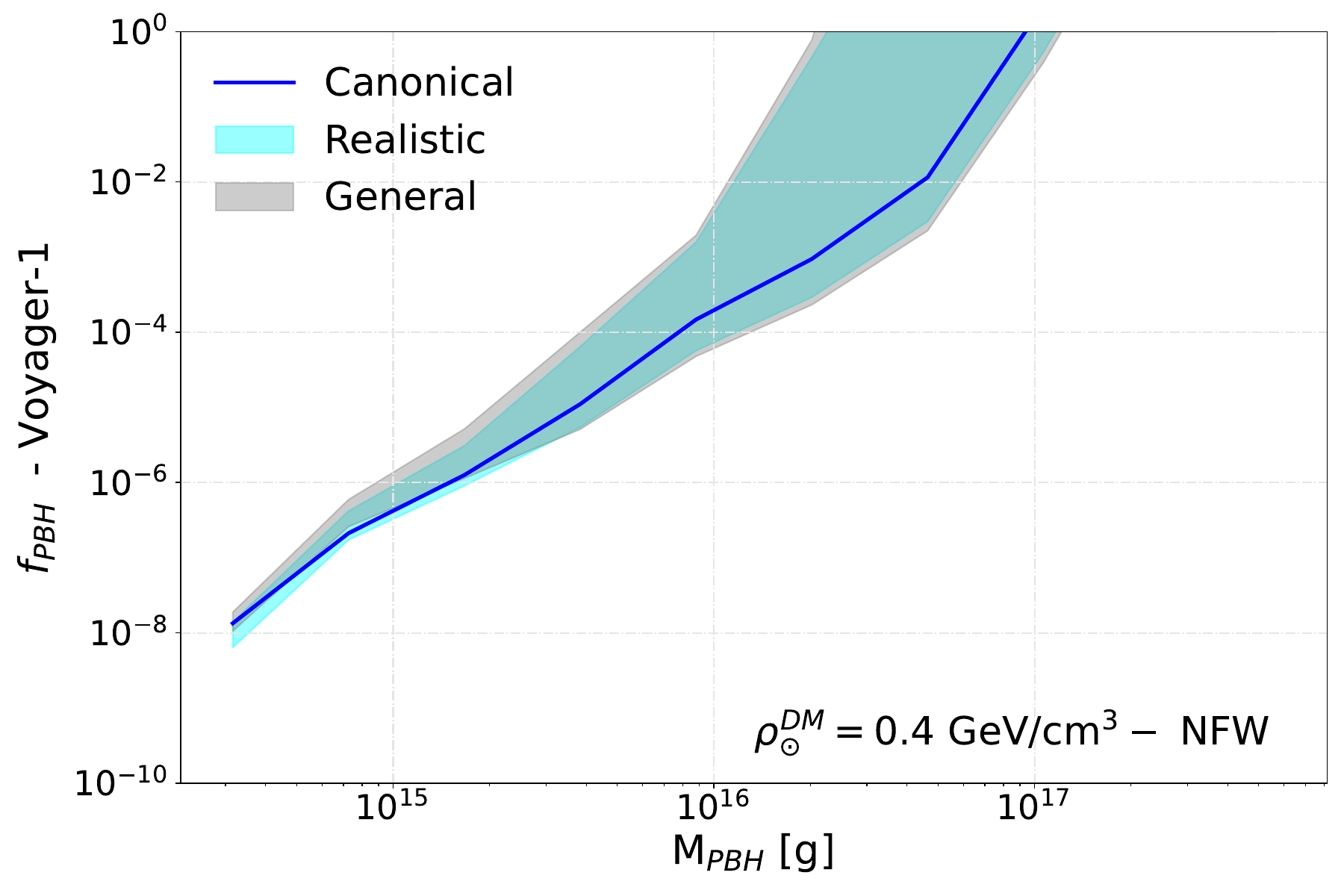}
    \includegraphics[width=0.46\linewidth]{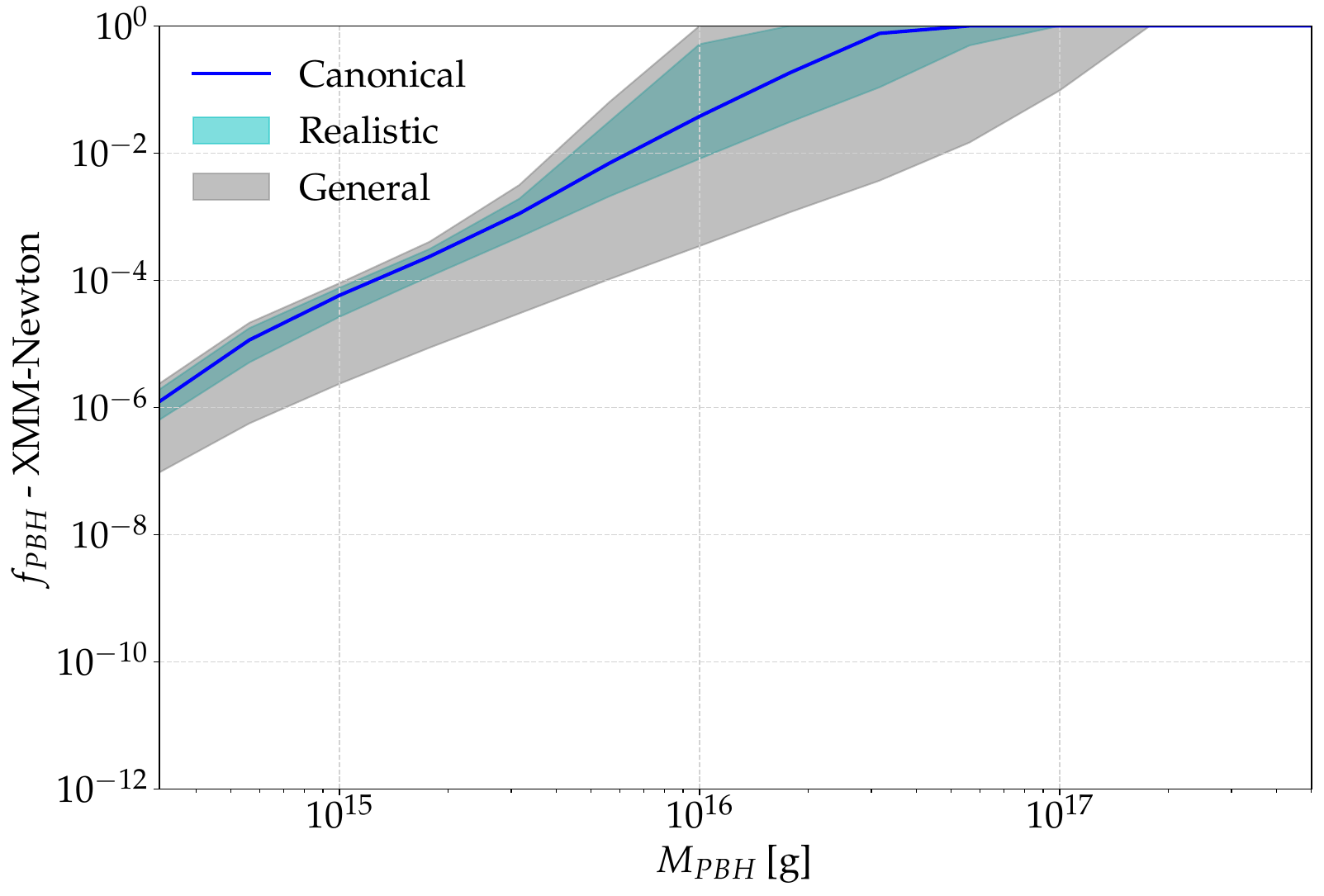}
    \includegraphics[width=0.465\linewidth]{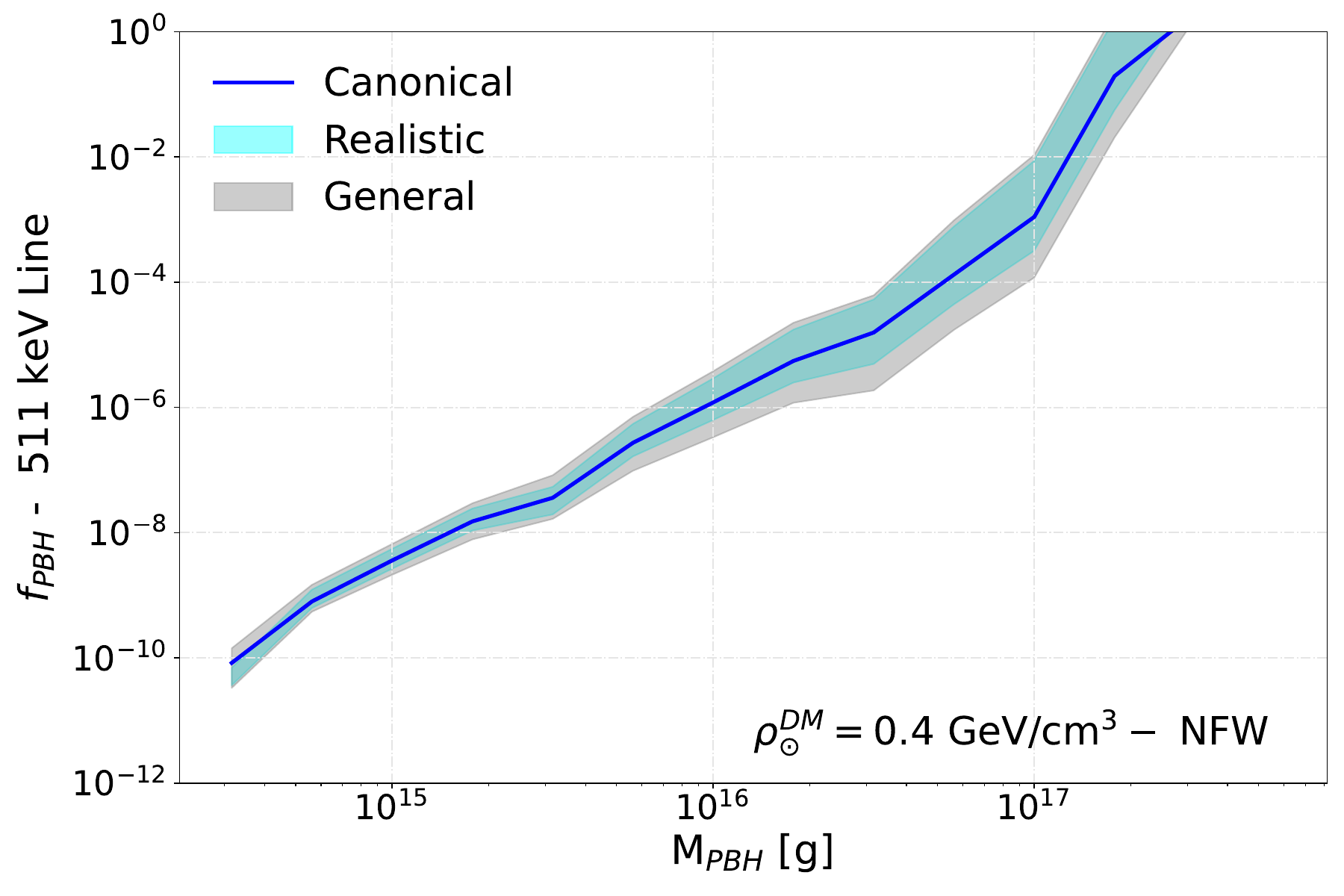}
    \caption{Uncertainties in the limits we derive using {\sc Voyager 1} (upper left panel), {\sc Xmm-Newton} (upper right panel) and the 511 keV line reported by {\sc Integral} (lower panel). The solid blue lines correspond to the limits using our fiducial propagation model. The blue bands show how our limits are impacted when varying the Alfvén speed $V_A$ and the halo height $H$ within their $3\sigma$ uncertainty. The grey bands correspond to a more conservative scenario where we vary $V_A$ between 0 and 40 km/s and $H$ between 3 and 16 kpc.}
    \label{fig:UncertsLimits}
\end{figure*}

Our $511$~keV bound, which we weaken by a factor of two to account for systematic uncertainties in the data (as mentioned when discussing the calculation of the $511$~keV line in Sec.~\ref{sec:propagation}), is shown in red in Fig.~\ref{fig:moneyplot}, where we compare with the limit reported in Ref.~\cite{Laha:2019ssq} (red dashed line). They used the rate of positron injection needed to explain the total $511$~keV flux from {\sc Integral} in the bulge. Given that the high-longitude measurements of the  $511$~keV line emission are the most constraining measurements, the use of longitudinal profile lead to more stringent results compared to using the bulge emission~\cite{DelaTorreLuque:2023cef}. In addition, the authors include only the emission from a NFW DM profile within the inner $3$~kpc from the Galactic Center and did not model positron propagation. Therefore our 511 keV bound appears to be more stringent than the one of Ref.~\cite{Laha:2019ssq}. In a recent paper~\cite{DelaTorreLuque:2024zsr}, we show that using in-flight positron annihilation emission can improve the limits on feebly interacting particles (whose $e^{\pm}$ emission is also concentrated at tens of MeV and follow a similar spatial morphology) with respect to the $511$~keV constraints, given that measurements of the diffuse $\gamma$-ray emission above a few MeV have a reduced systematic uncertainty and more reliable background models can be used.

Finally, we mention that a bound can be derived by requiring that the amount of heating of the intergalactic medium from PBH evaporation by 21-cm observations of the {\sc Edges} experiment~\cite{Mittal:2021egv}. All in all, our limit from the longitude profile of the $511$~keV line is competitive with the {\sc Edges} limit below $M_\textrm{PBH} \simeq 10^{16}$ g and becomes the most stringent limit to date for PBH masses between $10^{16}$ and $2\times 10^{17}$ g, for this propagation setup. However, it should be noted that the robustness of the {\sc Edges} experiment has been severely challenged since see e.g. Refs.~\cite{Hills:2018vyr,Bradley:2018eev,Singh:2021mxo}, hence we omit it from Fig.~\ref{fig:moneyplot}.

In Fig.~\ref{fig:moneyplot}, we assumed PBHs to be Schwarzschild BHs with a monochromatic mass distribution. The main reason is because non-rotating and monochromatic PBHs represent the most conservative case. However, if their mass distribution were instead log-normal, as in Eq.~\eqref{eq:lognorm}, there would be a low-mass PBH population that contributes to most of the flux of evaporated $e^\pm$ and photons, leading to a strengthening of the limits. Actually, for increasing values of the standard deviation $\sigma$ of the distribution, the low-mass population increases and therefore the limits become more and more stringent. Alternatively, if PBHs were Kerr BHs, they would produce more particles at high energies, ending up with a strengthening of the limits as well. Fig.~\ref{fig:SomeLimits} illustrates the impact of the choice of mass and spin distributions on the limits on $f_\textrm{PBH}$. Additionally, we have tested the impact of using different DM density distributions, comparing cuspier DM distributions than the NFW (i.e. slope of $\gamma>1$) with cored profiles, like a Einasto~\cite{Merritt_2006, Navarro_2009} or a Burkert profile~\cite{Burkert_1995}, finding that our limits are not expected to vary by more than a factor of $2$ (See Appendix~\ref{sec:App:XrayProfs}).

Finally, in Fig.~\ref{fig:UncertsLimits} we report the uncertainties on the limits we derive in this work, showing the impact of the choice of the propagation model, by using the propagation scenarios explained above. The blue bands (labeled `realistic') correspond to the variation of $V_A$ and $H$ up to their 3$\sigma$ uncertainties. For the lower side of the band we use $V_A = 20$ km/s and $H = 12$ kpc, while for the upper side we use $V_A = 7$ km/s and $H = 4$ kpc. Then the gray bands in Fig.~\ref{fig:UncertsLimits} (labeled `general') represent more conservative uncertainties, where for the lower side we adopt $V_A = 40$ km/s and $H = 16$ kpc, and for the upper side $V_A = 0$ km/s and $H = 3$ kpc. In general, these variations may affect our limits by up to an order of magnitude or more. In the case of the $511$~keV signals, we observe that the uncertainty bands are in general smaller. This is due to the fact that the morphology of the predicted $511$~keV line does not change significantly for different values of reacceleration (at high longitudes, where the main constraints come from, reacceleration does not appreciably change the emission, which is spatially very flat in any case). In the case of {\sc Voyager 1} and X-ray constraints, we observe that the uncertainty band typically broadens at higher PBH masses. The reason is that, high mass PBHs inject lower energy $e^{\pm}$, which reacceleration affects much more. For example, in the no reacceleration case the emission from PBHs of mass higher than a few times $10^{16}$~g lies below the {\sc Voyager 1} data points (therefore, no constraint can be set). Additionally, we note that uncertainties from the choice of DM distribution will have very little effect in the constraints from {\sc Voyager 1} and the $511$~keV line, while the limits from {\sc Xmm-Newton} can be significantly affected, given that the most constraining X-ray data is that coming from the inner regions of the Galaxy, which is where our predictions are more affected by uncertainties in the DM distribution.

\section{Conclusion}
\label{sec:conclusions}

In this work, we have conducted a thorough analysis of signals emanating from PBH evaporation thereby refining constraints on their role as DM candidates. Our study leverages observations from the {\sc Integral/Spi}, {\sc Voyager 1}, and {\sc Xmm-Newton}, integrating a comprehensive CR transport model that encapsulates reacceleration and diffusion within the Milky Way. By numerically solving the diffusion equations and employing current CR propagation frameworks, we have honed the limits on PBHs as DM, particularly for those with masses around $10^{16}$ g.

Our findings indicate that the limits derived from the $511$~keV line, $e^{\pm}$, and diffuse X-rays are significantly impacted by the assumptions regarding PBH mass, spin distributions and propagation models. They complement each other and significantly probe the parameter space available for PBHs as DM.
The most compelling result comes from the $511$~keV line emission at the Galactic disk (specifically, the high-longitude measurements of the longitudinal profile of the signal), where our analysis, assuming a NFW DM profile, yields the most stringent limits to date for PBH masses between $10^{16}$ and $2\times 10^{17}$~g. This bound is further corroborated by the heating constraints of the intergalactic medium from PBH evaporation as observed by the {\sc Edges} experiment. We additionally remark that our limits are conservatively derived without including backgrounds, that are expected, in all the studied cases, to be dominant. 
We note, however, that the X-ray constraints  produced from inverse Compton emission, are stronger than those from the $511$~keV line for optimistic diffusion parameters, as shown in Fig.~\ref{fig:UncertsLimits}, and also in the case of a cuspier DM distribution than the benchmark NFW.

Uncertainties in our limits, depicted in our analysis, underscore the sensitivity of our results to the choice of propagation model parameters, such as the Alfvén speed and halo height. Moreover, considering alternative mass distributions, like a log-normal distribution, or the inclusion of Kerr BHs, would lead to even more stringent limits due to the increased flux of evaporated particles and photons.
We note that the low mass part of the asteroid-mass gap is currently well probed and PBHs can constitute a significant fraction of the DM only in the gap between $10^{18}-10^{21}$~g in the monochromatic mass case and $\sim10^{19}-10^{21}$~g in the case where more realistic log-normal distribution is adopted for their mass distribution.

In conclusion, our comprehensive approach to analysing CR signals from PBH evaporation has not only refined current astrophysical constraints on PBHs as DM candidates but also highlighted the critical influence of propagation models and PBH distributions on these limits. Our work paves the way for future studies to further explore the intriguing possibility of PBHs constituting the elusive DM in our universe and also studying late-forming evaporating PBHs.

\acknowledgments
We acknowledge useful discussions with Alexandre Arbey, Joe Silk and Marco Cirelli.
SB is supported by the STFC under grant ST/X000753/1. PDL is supported by the Juan de la Cierva JDC2022-048916-I grant, funded by MCIU/AEI/10.13039/501100011033 European Union ``NextGenerationEU"/PRTR. The work of PDL is also supported by the grants PID2021-125331NB-I00 and CEX2020-001007-S, both funded by MCIN/AEI/10.13039/501100011033 and by ``ERDF A way of making Europe''. PDL also acknowledges the MultiDark Network, ref. RED2022-134411-T. This project used computing resources from the Swedish National Infrastructure for Computing (SNIC) under project Nos. 2021/3-42, 2021/6-326, 2021-1-24 and 2022/3-27 partially funded by the Swedish Research Council through grant no. 2018-05973.

\appendix

\section{Evaporation spectra from BlackHawk}
\label{appx:evap}

\begin{figure*}[!ht]
    \centering
    \includegraphics[width=0.49\linewidth]{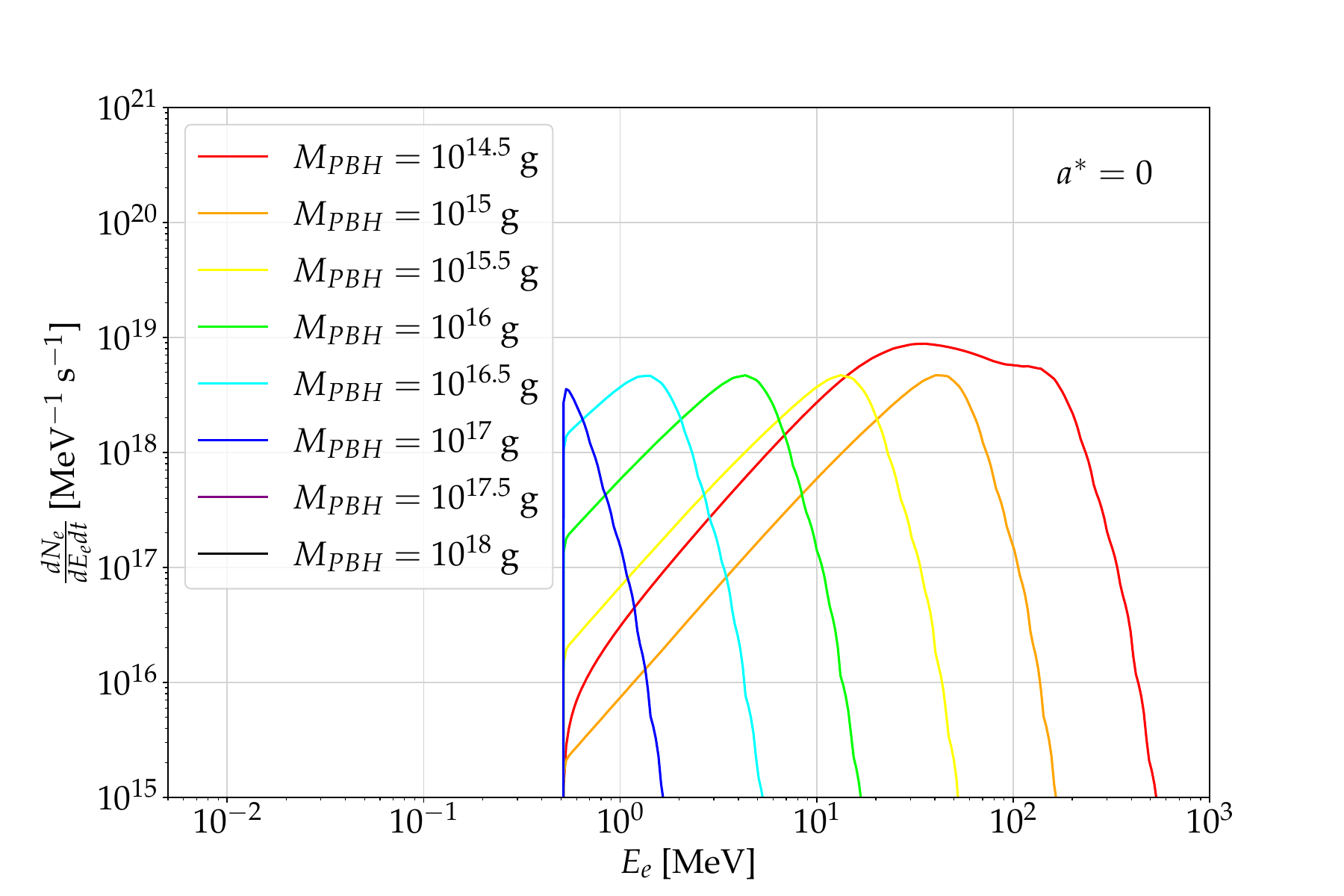}
    \includegraphics[width=0.49\linewidth]{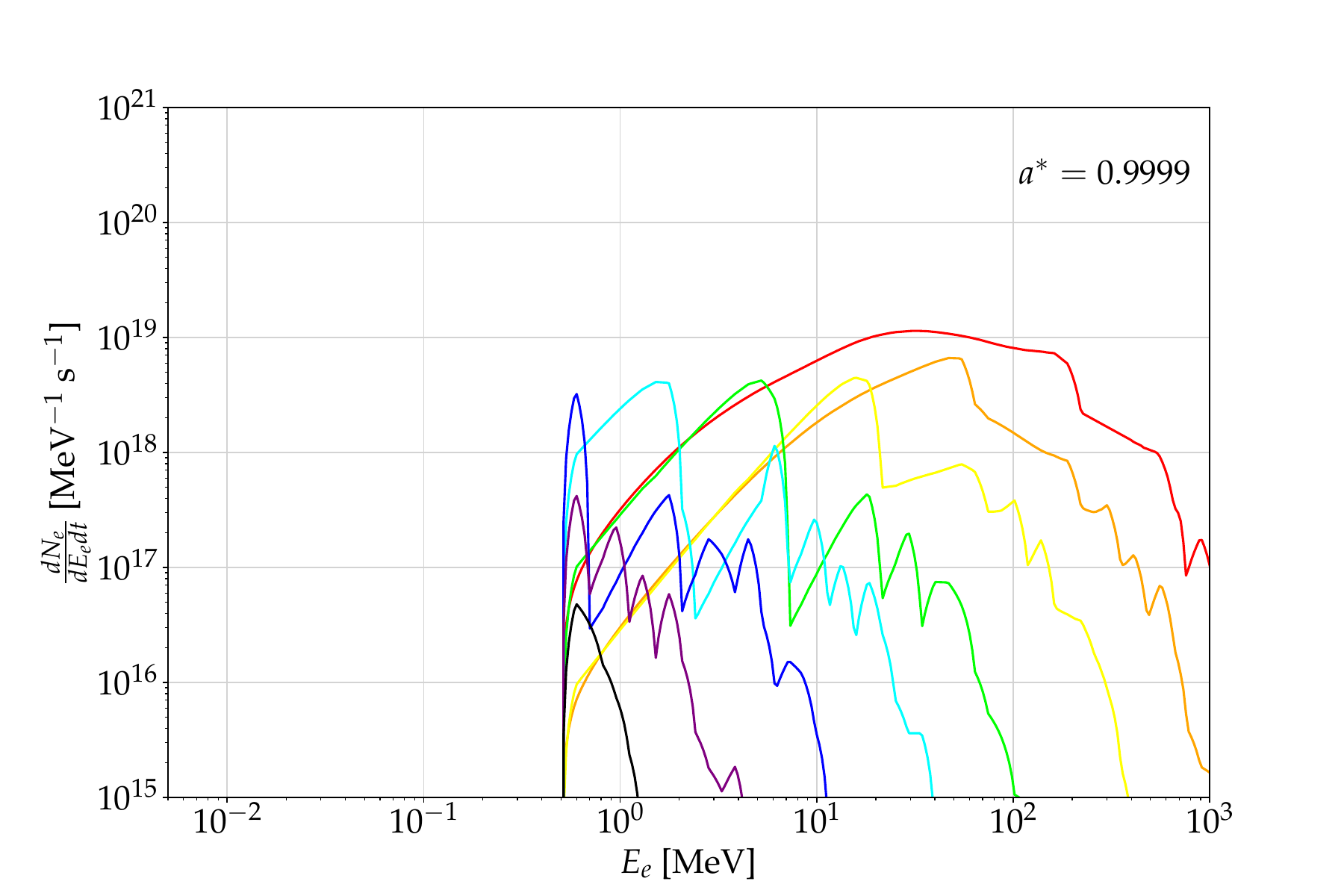}
    \cprotect\caption{Total spectra of $e^\pm$ from the evaporation of a single BH with spin of $a^\star = 0$ (left panel) and $a^\star = 0.9999$ (right panel), for the following BH masses:  $M = 10^{14.5}$ g (red), $10^{15}$ g (orange), $10^{15.5}$ g (yellow), $10^{16}$ g (lime), $10^{16.5}$ g (cyan), $10^{17}$ g (blue), $10^{17.5}$ g (purple) and $10^{18}$ g (black). Note that the emission from the heaviest two masses listed is too suppressed to be displayed in the left panel. Output of \verb|BlackHawk+Hazma|.}
    \label{fig:evapspec1}
    \bigskip
    \includegraphics[width=0.49\linewidth]{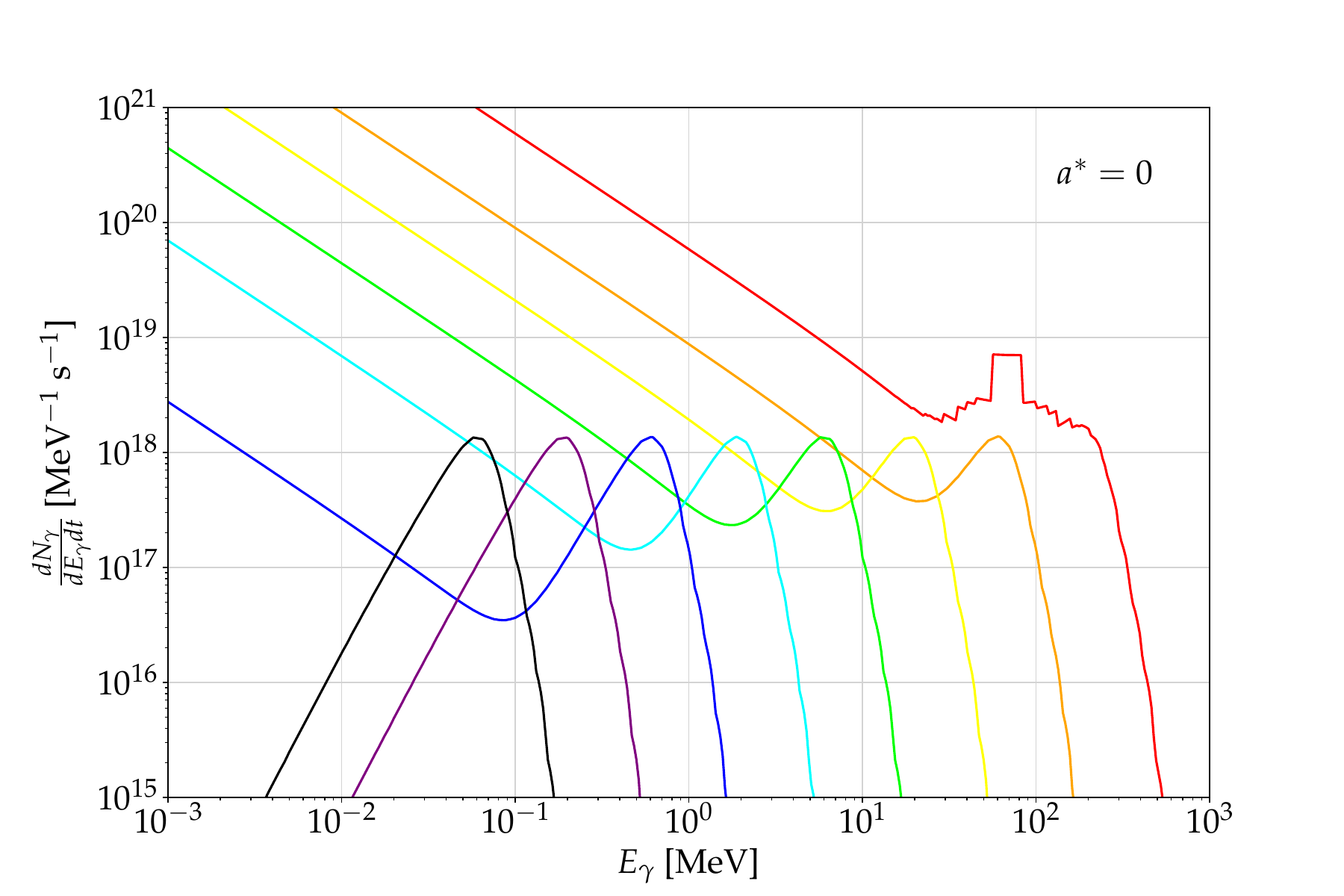}
    \includegraphics[width=0.49\linewidth]{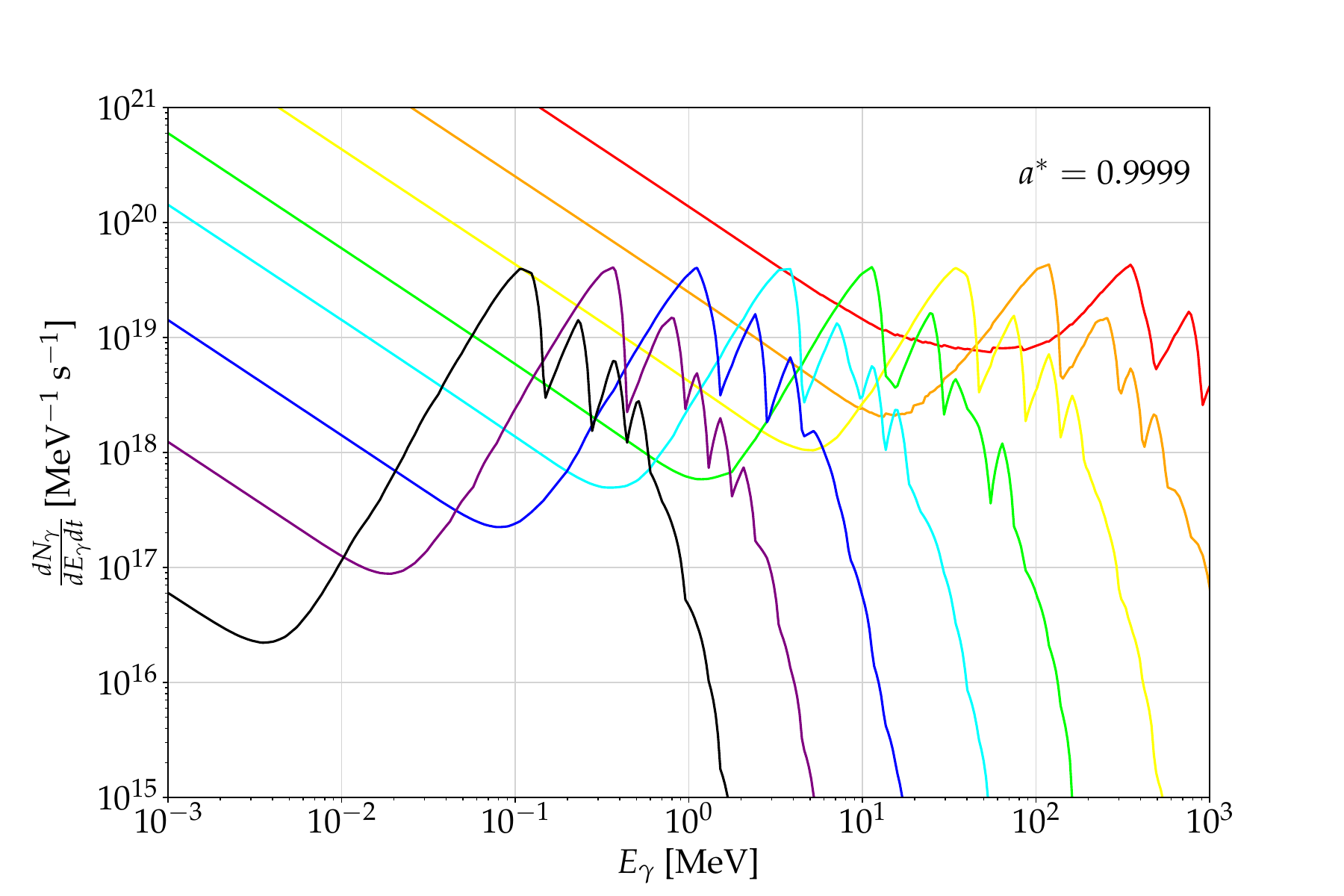}
    \cprotect\caption{Spectra of secondary $\gamma$ from the evaporation of a single BH with spin of $a^\star = 0$ (left panel) and $a^\star = 0.9999$ (right panel), for the following BH masses:  $M = 10^{14.5}$ g (red), $10^{15}$ g (orange), $10^{15.5}$ g (yellow), $10^{16}$ g (lime), $10^{16.5}$ g (cyan), $10^{17}$ g (blue), $10^{17.5}$ g (purple) and $10^{18}$ g (black). Output of \verb|BlackHawk+Hazma|. Same color scheme as in Fig.~\ref{fig:evapspec1}.}
    \label{fig:evapspec2}
\end{figure*}

In this appendix we discuss the spectra of secondary $e^\pm$ and $\gamma$ from the evaporation of a single BH. These are obtained using \verb|BlackHawk|, which deals with the evaporation of at most $\gamma$, $\nu_{e,\mu,\tau}$, $e^\pm$, $\mu^\pm$ and $\pi^{0,\pm}$ in the PBH mass range we consider in our study. For $M_\textrm{PBH} = M_\textrm{min}$ all of these particles are produced, whereas for $M_\textrm{PBH} = 10^{18}$ g only $\gamma$, $\nu_{e,\mu,\tau}$, $e^\pm$ are evaporated, recalling Eq.~\eqref{eq:BHtemp}. Then, $\mu^\pm$ and $\pi^{0,\pm}$ decay after being evaporated, in turn producing more $\gamma$, $\nu_{e,\mu,\tau}$, $e^\pm$. Moreover, all the charged particles we mentioned can emit final state radiations, increasing the count of $\gamma$ that originates from PBH evaporation as well. To compute the spectra of particles coming from these processes, we use the version of \verb|Hazma| integrated in \verb|BlackHawk|.
In Figs.~\ref{fig:evapspec1} and~\ref{fig:evapspec2} we show the total spectra of $e^\pm$ and $\gamma$ from BH evaporation for different BH masses and for $a^\star = 0$ (0.9999) in the left (right) panel. The spectrum of $e^\pm$ solely come from their emission from the PBH for $M_\textrm{PBH} \gtrsim 10^{14}$ g, however for lower masses, $\mu^\pm$ and $\pi^\pm$ start to be produced and their decay into $e^\pm$ contribute to the low-energy bump in the $e^\pm$ spectrum at $M_\textrm{PBH} = 10^{14.5}$ g. This also contributes to the bump feature we see in the $\gamma$ spectrum shown in the left panel of Fig.~\ref{fig:evapspec2}, although the low sampling in energy is a numerical artifact, probably due to the way that \verb|BlackHawk| interpolates the hadronization tables from \verb|Hazma|. In any case, this effect is present only for the lowest PBH masses which represent only the edge of the whole PBH mass range we considered in our study. Also, in the $\gamma$ spectra and for masses where $e^\pm$ start to be produced efficiently, the low-energy ramp corresponds to final state radiations, for which $dN_\textrm{FSR}/dE \propto 1/E$. Finally, we can witness that Kerr BHs emit more particles at higher energies than Schwarzschild ones.

\section{Impact of different DM profiles}
\label{sec:App:XrayProfs}
The DM density distribution affects the flux of particles produced by PBH evaporation. We expect the largest differences to be found around the center of the Galaxy, where the uncertainty in the DM distribution is the greatest. Therefore, to test the uncertainties in our limits related to the DM profile, we have computed the X-ray flux for a PBH of mass $M_\textrm{PBH} = 10^{16}$~g for an Einasto, cNFW ($\gamma = 1.25$), Moore ($\gamma = 1.5$) and Isothermal profile. As can be seen from Fig.~\ref{fig:Profs}, the largest differences are smaller than a factor of $2$. This means that uncertainties due to $e^{\pm}$ diffusion dominate those from the DM profile by far.

\begin{figure}[h!]
    \centering
    \includegraphics[width=0.9\linewidth]{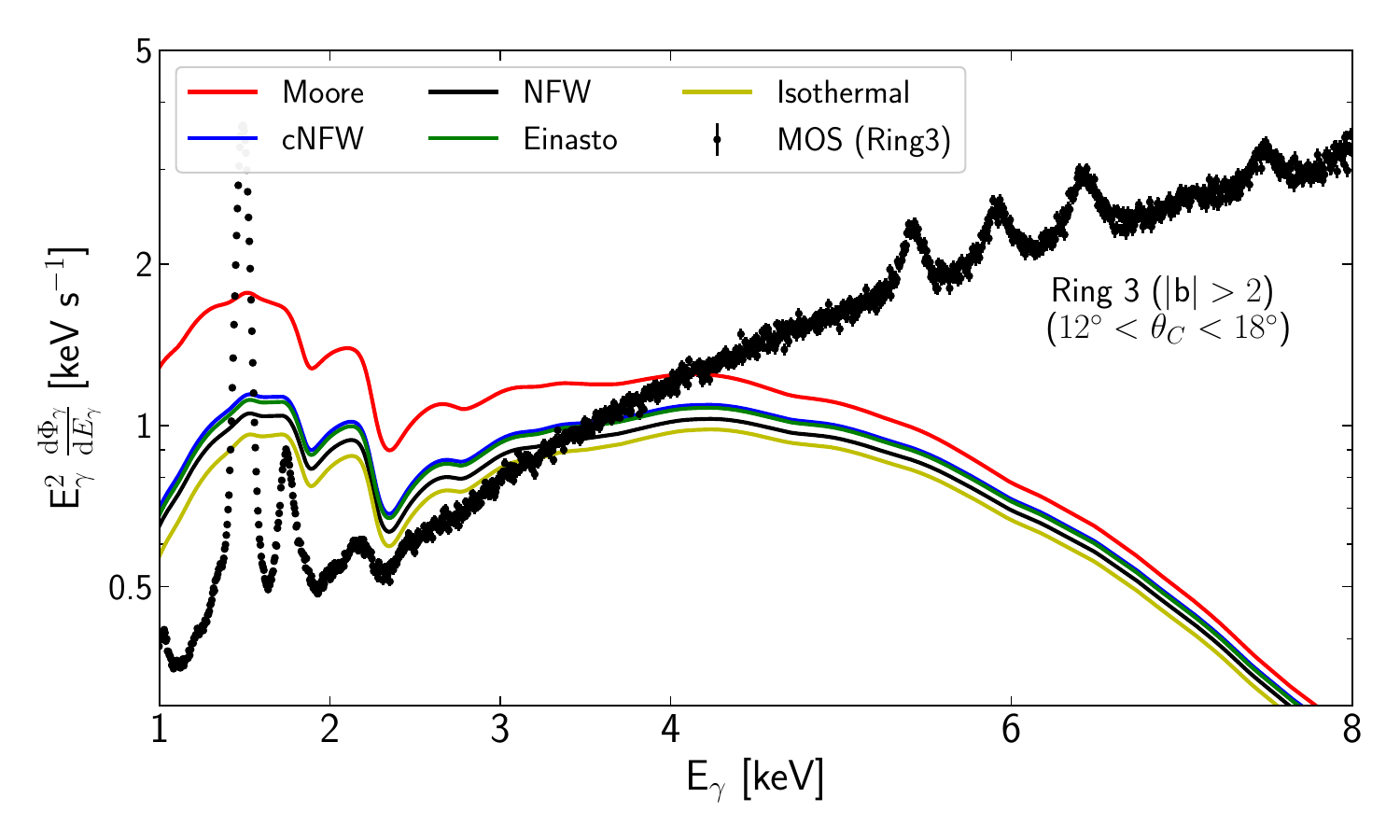}
    \caption{$X$-ray predicted flux from different DM profiles compared to XMM-data from a region close to the Galactic Center.}
    \label{fig:Profs}
\end{figure}

For the Voyager-1 the uncertainty associated to the choice of DM profile is even lower, given that these constraints are based local observations, very far away from the Galactic Center. Similarly, given that the constraint from the $511$~keV line is mainly obtained from the high longitude data points, this limit is not expected to change significantly (see Fig.~\ref{fig:511_profs}).


\clearpage
\onecolumngrid   

\section*{Erratum version}
\subsection*{Abstract}
We correct the results obtained using the diffuse $X$-ray emission from {\sc Xmm-Newton}, in light of new results from Ref.~\cite{Balaji:2025afr}, that demonstrated that the dataset employed was misevaluated. We also update our calculation of the $511$~keV emission from evaporating PBHs, which leads to slightly more conservative constraints as well.
\subsection*{Main text}
\setcounter{figure}{0}
\renewcommand{\thefigure}{S\arabic{figure}}
\renewcommand{\thesection}{S\Roman{section}}


\title{Erratum: Refining Galactic primordial black hole evaporation constraints}

\author{Pedro De la Torre Luque}
\email{pedro.delatorre@uam.es}
\affiliation{Departamento de F\'isica Te\'orica, M-15, Universidad Aut\'onoma de Madrid, E-28049 Madrid, Spain}
\affiliation{
    Instituto de F\'isica Te\'orica, IFT UAM-CSIC,
    Departamento de F\'isica Te\'orica,
    Universidad Aut\'onoma de Madrid,
    ES-28049 Madrid,
    Spain
}

\author{Jordan Koechler}
\email{jordan.koechler@gmail.com}
\affiliation{
    Istituto Nazionale di Fisica Nucleare, Sezione di Torino, Via P. Giuria 1, 10125 Torino, Italy
}

\author{Shyam Balaji}
\email{shyam.balaji@kcl.ac.uk}
\affiliation{
    Physics Department,
    King's College London,
    Strand, London, WC2R 2LS,
    United Kingdom
}

\date{\formatdate{\day}{\month}{\year}, \currenttime}


        


Ref.~\cite{Balaji:2025afr}, alongside with private communication with the authors of Ref.~\cite{XMM}, suggest that recent bounds derived using {\sc Xmm-Newton} data need to be revised. This arises from a misinterpretation regarding the solid angle of the observations with respect to the geometrical factor of a given ROI (which are $30$ co-centric rings centered on the GC), this has also affected other recent DM indirect searches. In Ref.~\cite{XMM}, they provide data that are normalized by the exposure-weighted average solid angle, which is orders of magnitude smaller than the geometric one. Taking for example Ring 3 ($12$°-$18$°), which was relevant to our work, the exposure-weighted average solid angle is $3.463\times10^{-5}$ sr, while the geometric angle is $0.156$ sr, which is more than $4$ orders of magnitude higher.

In our work, the prediction of the $X$-ray flux from PBH evaporation collected by {\sc Xmm-Newton} in the Ring 3 took into account the geometric angle, It was therefore dramatically overestimated (See Fig.~\ref{fig:XMMComp}).

Moreover, we take the opportunity to update our estimations for the $511$~keV line emission. In Ref.~\cite{laTorreLuquePedro:2024est} we noticed that our calculation of the $511$~keV flux over-counted the number of positrons contributing to positronium formation. This is due to the fact that we integrated the positron flux over all energies, thereby considering high-energy positrons that typically escape or annihilate in-flight before thermalizing, and therefore will not produce $511$~keV photons in the region studied. We have updated our calculation to account only for the thermalized positrons, following Refs.~\cite{laTorreLuquePedro:2024est, DelaTorreLuque:2023cef}. This correction weakens the PBH fraction limits at low PBH masses, while the effect is not appreciable at high PBH masses, around $\sim10^{17}$~g. The spatial morphology of the signals remains roughly unchanged, since they mostly inherit the source DM density profile adopted (See Fig.~\ref{fig:511Comp}).

In light of these new revisions, we provide the corrected figures (Figs.~\ref{fig:XMMComp}, \ref{fig:511Comp}, \ref{fig:SomeLimits} and~\ref{fig:moneyplot}) that refer to the results derived using the corrected {\sc Xmm-Newton} data and the updated estimation of the $511$~keV signal associated to PBH evaporation.

\begin{figure*}[ht!]
    \centering
    \includegraphics[width=0.49\linewidth]{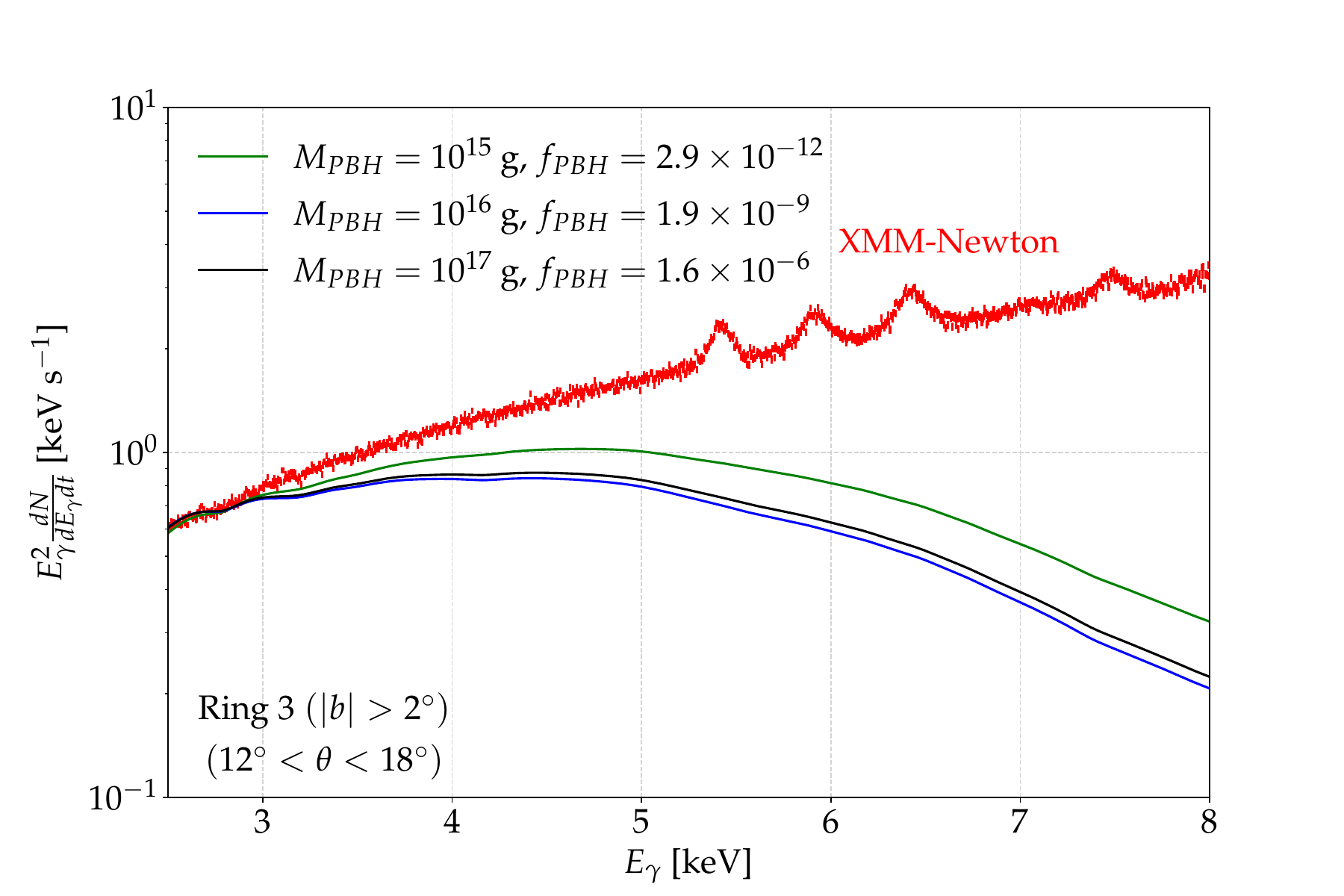}
    \includegraphics[width=0.49\linewidth]{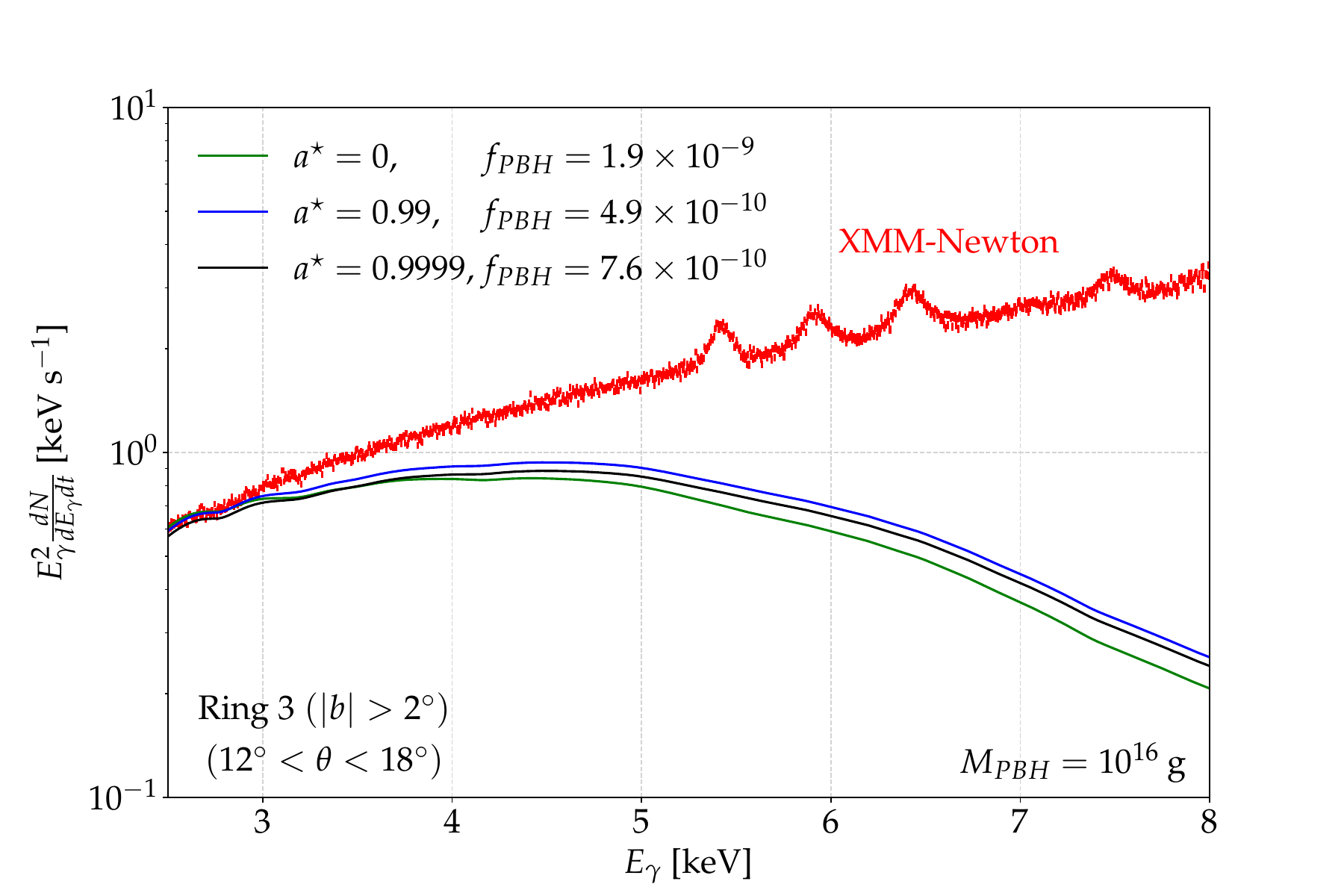}
    \includegraphics[width=0.49\linewidth]{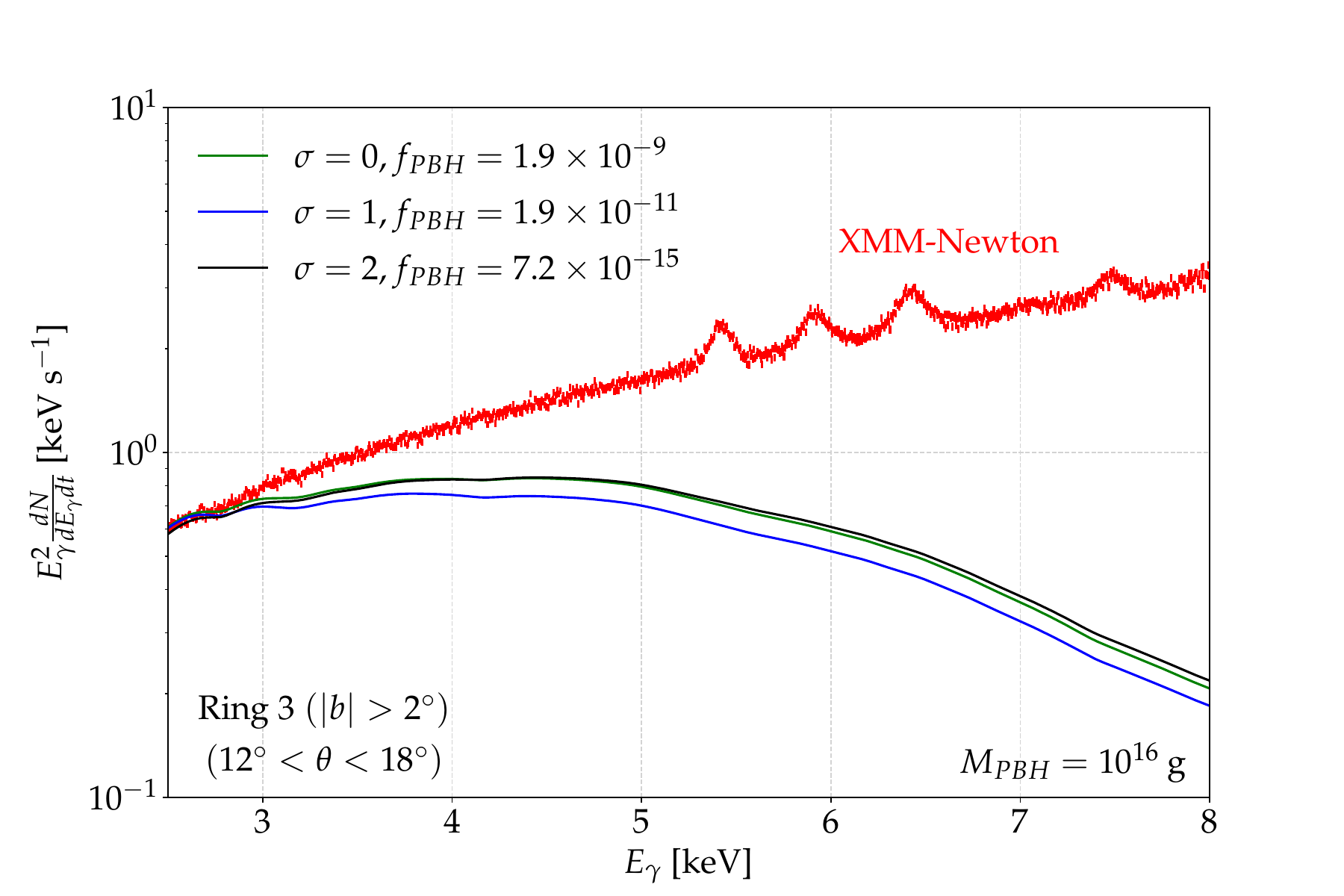}
    \caption{Comparison of the predicted DM-induced X-ray emission with diffuse X-ray data from {\sc Xmm-Newton} in a region close to the Galactic Center. We show the prediction is shown for different values of $M_\textrm{PBH}$ when the PBH mass distribution is monochromatic (upper left panel), of $a^\star$ (upper right panel) and $\sigma$ when the distribution is log-normal (lower panel).}
    \label{fig:XMMComp}
\end{figure*}

\begin{figure*}[ht!]
    \centering
    \includegraphics[width=0.49\linewidth]{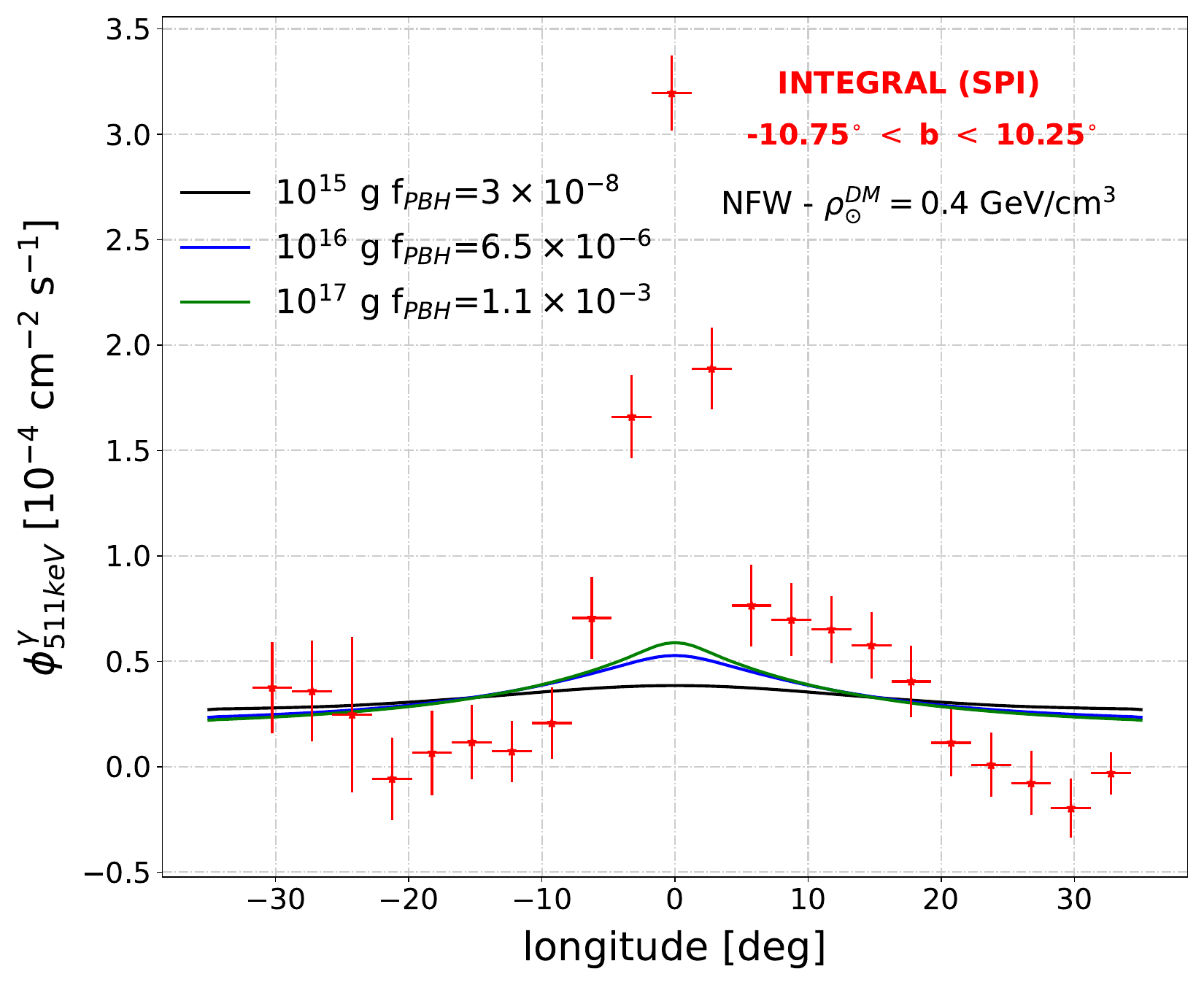}
    \includegraphics[width=0.49\linewidth]{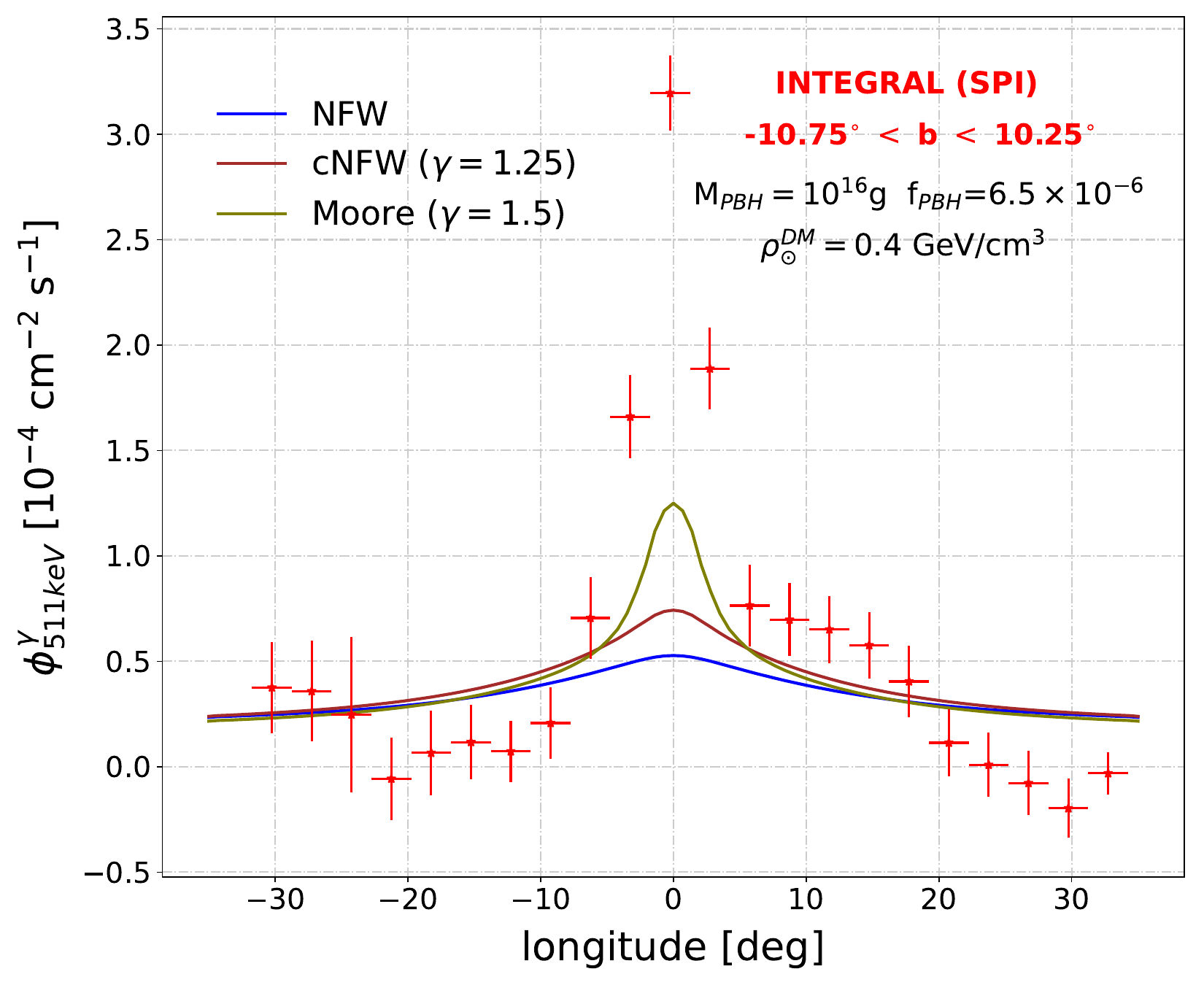}
    \caption{Comparison of the predicted 511 keV line longitudinal profile with SPI data. The left panel shows a comparison of the signal for different PBH masses following an NFW profile, while the right panel shows a comparison of the signals for a fixed mass of $10^{16}$~g following different popular DM density distributions.}
    \label{fig:511Comp}
\end{figure*}

\begin{figure*}[ht!]
    \centering
    \includegraphics[width=0.49\linewidth]{figures/PBH_XMM_limits_v4.pdf}
    \includegraphics[width=0.49\linewidth]{figures/PBH_XMM_limits_uncert_v4.pdf}
    \includegraphics[width=0.49\linewidth]{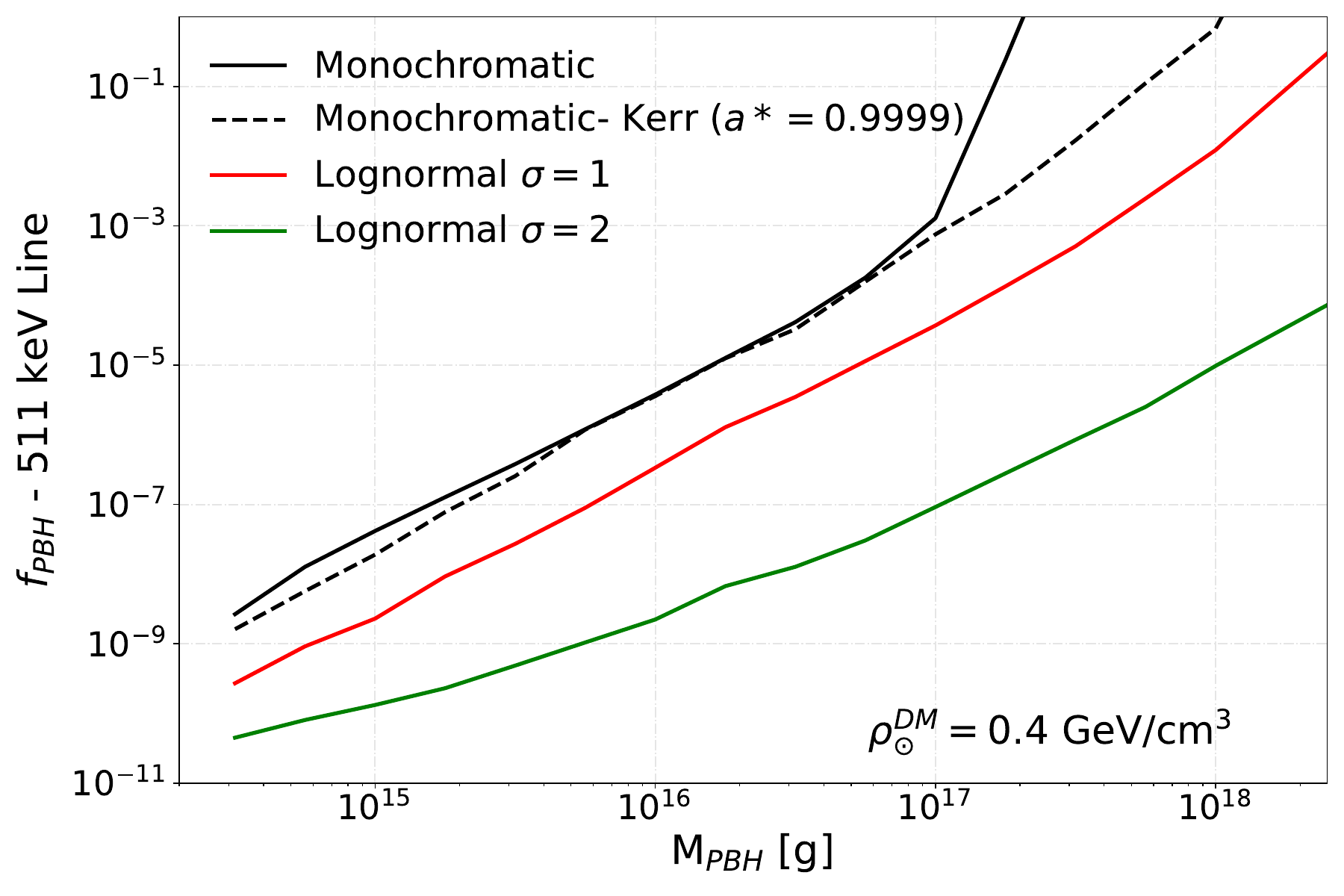}
    \includegraphics[width=0.49\linewidth]{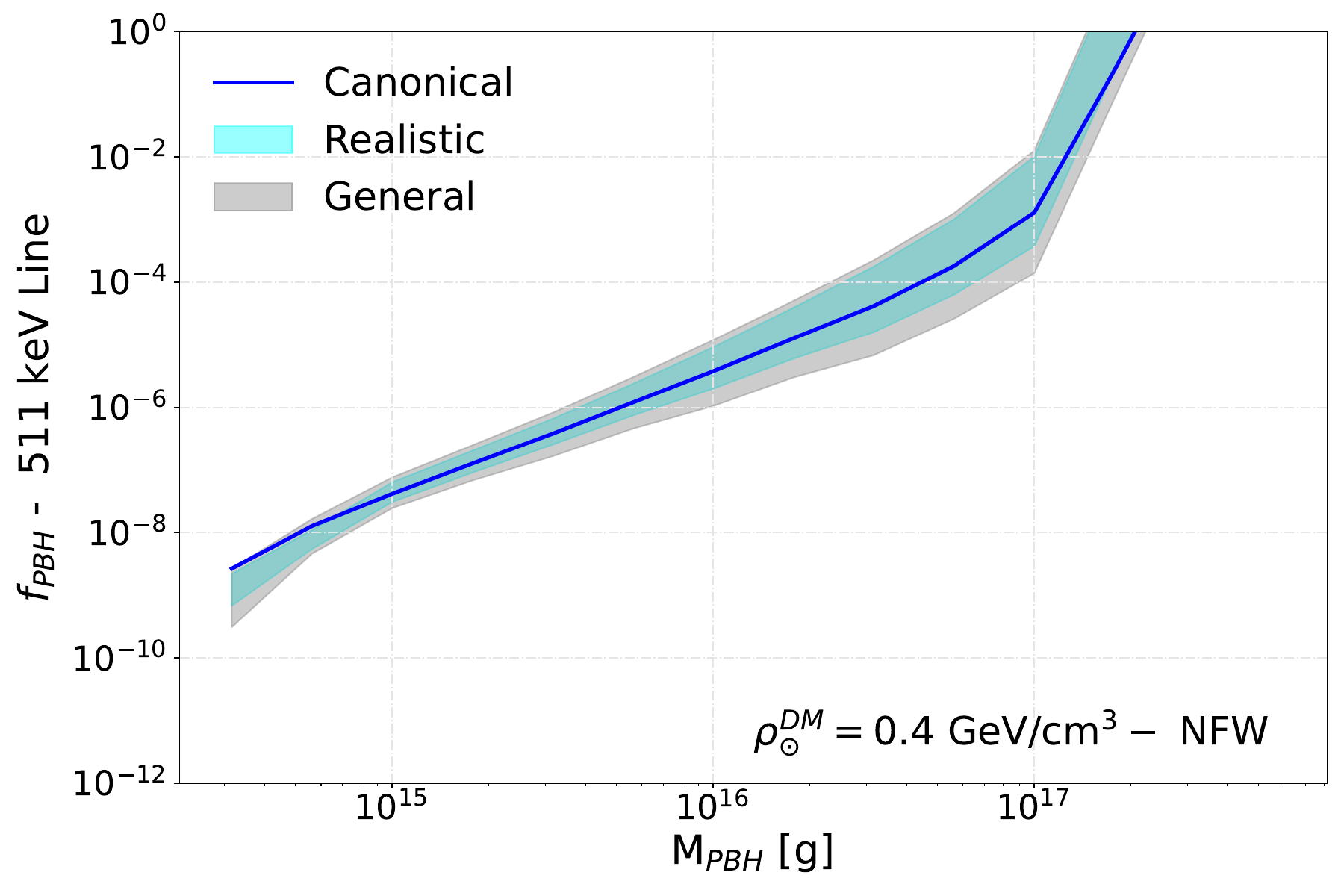}
    \caption{Top left Panel: Limits we derive using {\sc Xmm-Newton} for mass distributions with $\sigma = 0$ (monochromatic, solid black line), 1 (red line) and 2 (green line), as well as for the extreme assumption where all PBHs have a spin of $a^\star = 0.9999$ (dashed line). Top right panel: Uncertainties in the limits we derive using {\sc Xmm-Newton}. The solid blue lines correspond to the limits using our fiducial propagation model. The blue bands show how our limits are impacted when varying the Alfvén speed $V_A$ and the halo height $H$ within their $3\sigma$ uncertainty. The grey bands correspond to a more conservative scenario where we vary $V_A$ between 0 and 40 km/s and $H$ between 3 and 16 kpc. Bottom panels are similar to the top panels but for the calculation of the 511 keV line emission.}
    \label{fig:SomeLimits}
\end{figure*}

The conclusions of our papers remain unchanged, although the XMM-Newton constraints on $f_\text{PBH}$ weakened substantially. The $511$~keV constraint was corrected and remains one of the leading existing constraints, while the Voyager-1 constraint required no revision.

\begin{figure*}[ht]
    \centering
    \includegraphics[width=0.65\linewidth]{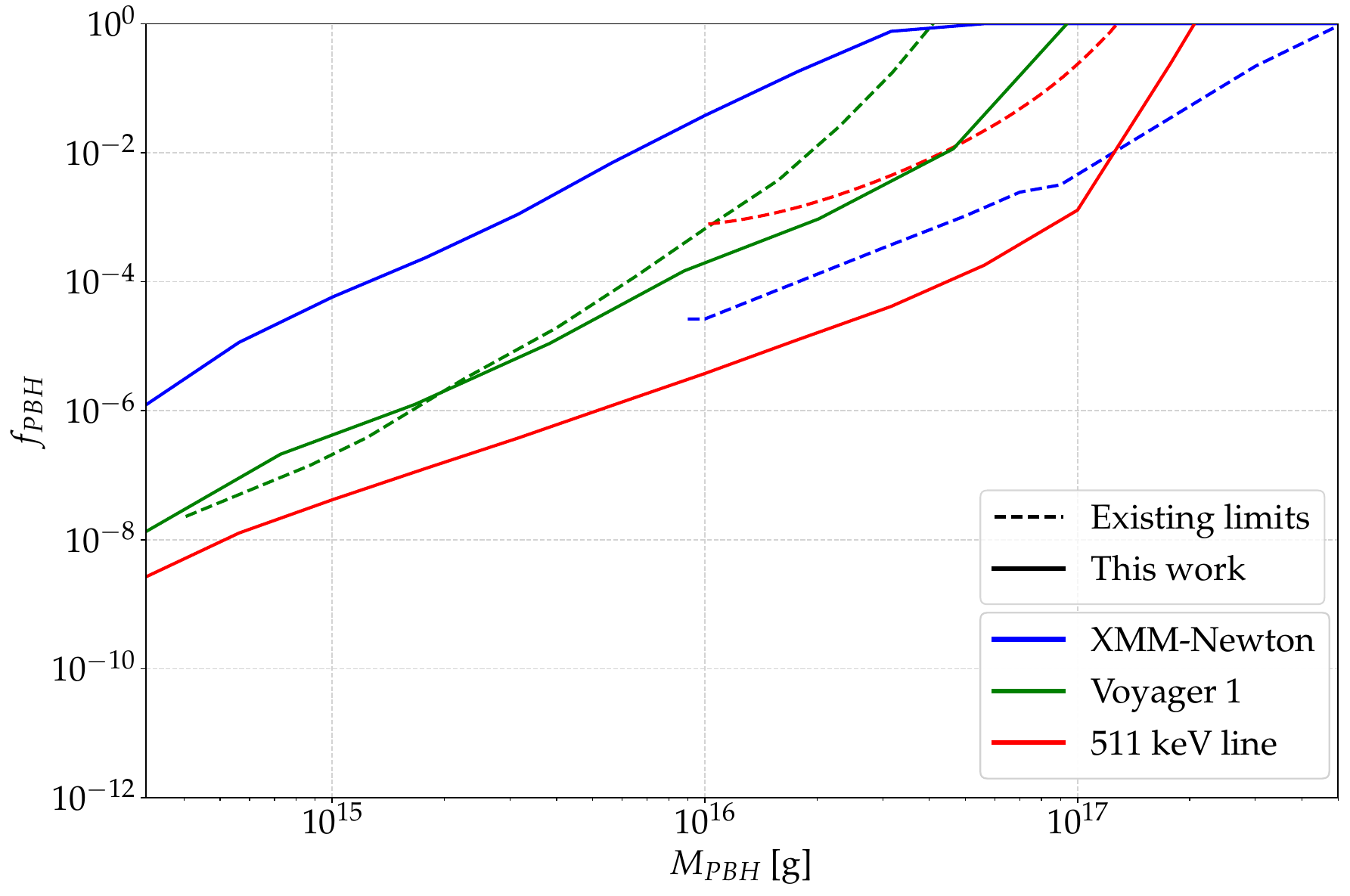}
    \cprotect\caption{Comparison of the 95\% confidence limits on $f_\textrm{PBH}$ derived in this work with other existing ones. The color of the lines represent the different probes used to set the constraints: green for the $e^\pm$ measurements from {\sc Voyager 1}, blue for X-ray diffuse observations from {\sc Xmm-Newton}, red for the 511 keV excess reported by {\sc Integral}. The two different line styles correspond to either the bounds derived either in this work (solid) or in the literature~\cite{Boudaud:2018hqb,Laha:2019ssq,Tan:2024nbx,Mittal:2021egv} (dashed) reported in \verb|PBHbounds|~\cite{kavanagh_2019}.}
    \label{fig:moneyplot}
\end{figure*}


\clearpage
\bibliographystyle{apsrev4-1}
\bibliography{references}

\end{document}